\documentclass[aps,twocolumn,groupedaddress,showpacs,showkeys,nofootinbib,amsfonts,eqsecnum,floatfix]{revtex4}

\usepackage{graphicx}
\usepackage{epsfig}
\usepackage{bm}

\newcommand{\diag}{{\textrm {diag}}}
\newcommand{\tr}{{\textrm {tr}}}

\newcommand{\Det}{{\textrm {Det}}}

\newcommand{\om}{{\widehat \omega}}

\def\slashchar#1{\setbox0=\hbox{$#1$}
   \dimen0=\wd0 \setbox1=\hbox{/} \dimen1=\wd1
   \ifdim\dimen0>\dimen1 \rlap{\hbox to \dimen0{\hfil/\hfil}} #1
   \else  \rlap{\hbox to \dimen1{\hfil$#1$\hfil}} / \fi}

\def\k{\slashchar{k}}

\begin{document}

\title{ Polyakov loop in chiral quark models at finite temperature }

\author{E. Meg\'{\i}as}
\email{emegias@ugr.es}

\author{E. \surname{Ruiz Arriola}}
\email{earriola@ugr.es}

\author{L.L. Salcedo}
\email{salcedo@ugr.es}

\affiliation{
Departamento de F\'{\i}sica At\'omica, Molecular y Nuclear,
Universidad de Granada,
E-18071 Granada, Spain
}

\date{\today}

\begin{abstract}
We describe how the inclusion of the gluonic Polyakov loop
incorporates large gauge invariance and drastically modifies finite
temperature calculations in chiral quark models after color neutral
states are singled out. This generates an effective theory of quarks
and Polyakov loops as basic degrees of freedom. We find a strong
suppression of finite temperature effects in hadronic observables
triggered by approximate triality conservation (Polyakov cooling), so
that while the center symmetry breaking is exponentially small with
the constituent quark mass, chiral symmetry restoration is
exponentially small with the pion mass.  To illustrate the point we
compute some low energy observables at finite temperature and show
that the finite temperature corrections to the low energy coefficients
are $N_c$ suppressed due to color average of the Polyakov loop. Our
analysis also shows how the phenomenology of chiral quark models at
finite temperature can be made compatible with the expectations of
chiral perturbation theory. The implications for the simultaneous
center symmetry breaking-chiral symmetry restoration phase transition
are also discussed.
\end{abstract}

\pacs{11.10.Wx 11.15.-q  11.10.Jj 12.38.Lg }

\keywords{finite temperature; heat kernel expansion; effective action;
gauge invariance; chiral quark models; chiral Lagrangian; Polyakov Loop}

\maketitle

\section{INTRODUCTION}
\label{sec:intro}

The general belief that QCD undergoes a phase transition to a
quark-gluon plasma phase at high temperature has triggered a lot of
activity both on the theoretical as well as on the experimental
side. The original argument put forward by Casher~\cite{Casher:1979vw}
suggesting that confinement implies dynamical chiral symmetry breaking
and hence that the chiral and deconfinement phase transitions take
place simultaneously at least at zero chemical potential, has been
pursued and so far confirmed in theoretical studies on the
lattice~\cite{Karsch:1998hr}. This also agrees with the
phenomenological determinations of the vacuum energy density in the
bag model, with an energy density difference between the Wigner and
Goldstone realizations of chiral symmetry. It has also been shown that
in the large $N_c$ limit with the temperature $T$ kept fixed, if a
chiral phase transition takes place it should be first
order~\cite{Gocksch:1982en}.

The coupling of QCD distinctive order parameters at finite temperature
to hadronic properties has been the subject of much attention over the
recent past \cite{Sannino:2002wb,Mocsy:2003qw,
Fukushima:2003fw,Fukushima:2003fm,Gocksch:1984yk,Meisinger:2003uf}
mainly in connection with theoretical expectations on the formation of
quark-gluon plasma and the onset of deconfinement. Indeed, even if
such a state of matter is produced in existing (RHIC, SPS
\cite{McLerran:2002jb,Heinz:2002gs}) and future (LHC) facilities,
the states which are detected are hadrons created in a hot
environment. Thus, it makes sense to study the properties of hadrons
in a medium which can undergo a confinement-deconfinement phase
transition. For heavy masses, quarks become static sources and there
is a general consensus that the order parameter can be taken to be the
Polyakov loop or thermal Wilson line \cite{KorthalsAltes:2003sv} where
the breaking of the center symmetry signals the onset of
deconfinement.  Dynamical light quarks, however, break explicitly the
center symmetry and no criterion for deconfinement has been
established yet \cite{Meyer:1983hm,Svetitsky:1986ye}. In QCD, there
has been an increasing interest in developing effective actions for
the Polyakov loop as a confinement-deconfinement order parameter
because of their relevance in describing the phase transition from
above the critical temperature~\cite{Dumitru:2001xa,Meisinger:2001cq,
Dumitru:2003hp,Meisinger:2003id}.

On the other hand, in a hot medium, one also expects that the
spontaneously broken chiral symmetry is restored at some critical
temperature. For this chiral phase transition the quark condensate is
commonly adopted as the relevant order parameter. The melting of the
chiral quark condensate has been observed on the
lattice~\cite{Karsch:1998hr}, is suggested by chiral perturbation
theory extrapolations~\cite{Gasser:1986vb,Gerber:1988tt} and is
numerically reproduced in chiral quark models
before~\cite{Bernard:1987ir,Christov:1991se} and after inclusion of pion
corrections~\cite{Florkowski:1996wf} (for a review see
e.g. Ref.~\cite{Oertel:2000jp}).

Where theory has most problems is precisely in the interesting
intermediate temperature regime around the phase transition, because
both the lightest Goldstone particles and the Polyakov loop degrees of
freedom should play a role, if they coexist. Up to now it is uncertain
how the corresponding states couple to each other from a fundamental
QCD viewpoint, hence some modeling is required. Based on previous
works~\cite{Meisinger:1995ih, Fukushima:2003fw,Gocksch:1984yk} and to
comply with chiral symmetry it seems natural to couple chiral quark
models and the Polyakov loop in a minimal way as an effective space
dependent color chemical potential. The work in
Ref.~\cite{Gocksch:1984yk} accounts for a crossover between the
restoration of chiral symmetry and the spontaneous breaking of the
center symmetry, reproducing qualitatively the features observed on
the lattice simulations \cite{Kaczmarek:2002mc} and which find a
natural explanation in terms of dimension two
condensates~\cite{Megias:2005ve}. In this regard we want to argue
below that the special role played by the gauge symmetry at finite
temperature actually requires this coupling, and elaborate on the
consequences of it when the quantum gluon effects are considered.

The organization of the paper is as follows. We review some facts on
large gauge symmetry at finite temperature in Sect.~\ref{sec:lgt}
which are put into the context of chiral quark models. Next, we
address the problem suffered by chiral quark models at finite
temperature in Section~\ref{sec:problem} where we argue that the
origin of the difficulty is related to a defective treatment of the
large gauge symmetry at finite tempe\-rature. Thus, to comply with
gauge invariance at finite temperature one should at least couple the
quarks to the $A_0 $ gluon field. We do this in Section
\ref{sec:polyakovcoupling}. This is equi\-valent to make the
replacement
\begin{eqnarray}
\partial_0 \to \partial_0 + i A_0   
\end{eqnarray} 
which corresponds to an $\vec{x}$-dependent chemical potential
coupling in the color fundamental representation. Obviously, this
coupling introduces a color source into the problem for a fixed $A_0$
field. In order to project onto the color neutral states we integrate
over the $A_0$ field, in a gauge invariant manner. In
Section~\ref{sec:oneloop} we describe the consequences of such a
coupling and projection in chiral quark models for a variety of
observables at the one quark loop approximation. Actually, as we will
show, there is an accidental $\mathbb{Z}(N_c)$ symmetry in the model
which gene\-rates a triality (super)selection rule at this level of
approximation, from which a strong thermal suppression, $ {\cal O}
(e^{-N_c M /T} )$ follows in the quenched approximation. This puts
some doubts on whether chiral quark models do predict a chiral phase
transition at realistic temperatures as we advanced in previous
communications~\cite{Megias:2004kc,Megias:2004gy}. Corrections beyond
one quark loop are discussed in Section~\ref{sec:corrections} where we
see that the suppression at low temperatures actually becomes ${\cal
O} (e^{-m_\pi /T})$, very much along the expectations of Chiral
Perturbation Theory (ChPT)~\cite{Gasser:1986vb}. Gluonic corrections
and local corrections in the Polyakov loop are also analyzed in this
section. In view of our discussions we illustrate in
Section~\ref{sec:unquenched} the situation with schematic dynamical
calculations involving quantum and local Polyakov loops in the
unquenched theory as compared to lattice studies. In
Section~\ref{sec:phase-tran} we extend these calculations to the
region around the phase transition. Finally, in
Section~\ref{sec:concl}, we summarize our points and draw our main
conclusions.

\section{Gauge invariance of chiral quark models at finite 
temperature and the Polyakov loop}
\label{sec:lgt}

In this section we review some relevant and naively very disparate
concepts of gauge symmetry at finite temperature,
Sect.~\ref{sec:large-gauge}, and the center symmetry in gluodynamics,
Sect.~\ref{sec:center}, as well as the standard chiral quark models,
Sect.~\ref{sec:cqmod}, in order to fix our notation for the rest of
the paper. Both subjects are well known on their own, although rarely
discussed simultaneously, and the reader familiar with any of them may
skip the corresponding subsections. Advancing the result of subsequent
discussions made in latter sections, the basic Polyakov Chiral Quark
Model is first introduced in Sect.~\ref{sec:pcqm}. The conflict
between both large gauge symmetry and chiral quark models is discussed
in Sect.~\ref{sec:problem}. The solution to the problem is elaborated
in Sect.~\ref{sec:polyakovcoupling} where the coupling of the Polyakov
loop to chiral quark models is motivated.

\subsection{Large gauge symmetry }
\label{sec:large-gauge}

One of the most striking features of a gauge theory like QCD at finite
temperatures is the non-perturbative manifestation of the non Abelian
gauge symmetry. Indeed, in the Matsubara formalism of quantum field
theory at finite temperature the space-time becomes a topological
cylinder: one introduces a compactified Euclidean imaginary
time~\cite{Landsman:1986uw} and evaluates the path integral subjecting
the fields to periodic or antiperiodic boundary conditions for bosons
and fermions respectively in the imaginary time interval $ \beta =1/T
$ where $T$ is the temperature. We use the Euclidean notation $x_4=i
x_0$ and $A_4(\vec x , x_4)=i A_0(\vec x , x_0)$.  Thus, only periodic
gauge transformations, $g(\vec x , x_4 ) = g(\vec x , x_4 + \beta ) $,
are acceptable since the quark and gluon fields are stable under these
transformations. In the Polyakov gauge $\partial_4 A_4 =0 $ with $A_4
$ a diagonal traceless $N_c \times N_c $ matrix, one has for the gauge
$SU(N_c)$ group element
\begin{eqnarray}
g (x_4 ) =  \diag(e^{i  2 \pi x_4n_j T })
\label{eq:1.1}
\end{eqnarray}
$(\sum_{j=1}^{N_c}n_j=0)$ the following gauge transformation on the
$A_4$ component of the gluon field
\begin{eqnarray}
A_4 \to A_4 +  2 \pi T \diag(n_j)\,.
\end{eqnarray} 
Thus, in this particular gauge, gauge invariance manifests as the
periodicity in the $A_4 $ gluon field. This property is crucial and
prevents from the very beginning from the use of a perturbative
expansion in the gluon field, $A_4 $, at finite temperature. This
large gauge symmetry\footnote{Technically speaking the transformations
(\ref{eq:1.1}) may not be large in the topological sense (i.e.,
homotopically non trivial). This depends on the topology of the
spatial manifold as well as on the gauge group
\cite{Garcia-Recio:2000gt}. They are topologically large within the
Polyakov gauge.} can be properly accounted for by considering the
Polyakov loop or untraced Wilson line as an independent degree of
freedom,
\begin{eqnarray}
\Omega ( x ) = {\cal T} \exp{ i \int_{x_4}^{x_4+1/T} d x_4^\prime
A_4 (\vec x , x_4^\prime) }
\end{eqnarray} 
where ${\cal T}$ indicates the Euclidean temporal ordering ope\-rator
and $A_4$ the gluon field. Under a general periodic gauge
transformation one gets
\begin{eqnarray}
\Omega (x) \to g(x) \Omega (x) g^\dagger (x) \,.
\end{eqnarray}
In the Polyakov gauge, which we assume from now on, $\Omega$ becomes
\begin{eqnarray}
\Omega(\vec{x})  = e^{ i A_4 (\vec x) /T }
\end{eqnarray} 
and so it is invariant under the set of transformations
(\ref{eq:1.1}). The failure of perturbation theory at finite
temperature in a gauge theory has generated lot of discussion in the
past mainly in connection with topological aspects, Chern-Simons terms,
anomalies, etc. In the case of the topolo\-gical Chern-Simons term
radiatively induced by fermions in $2+1$ dimensions
\cite{Redlich:1984kn} it was puzzling to find, in the perturbative
treatment, that the Chern-Simons quantization condition
\cite{Deser:1981wh} was violated at finite temperature
\cite{Pisarski:1986gq,Babu:1987rs}. It was subsequently shown that,
within a non perturbative treatment, no contradiction arises
\cite{Dunne:1996yb}. In \cite{Salcedo:1998sv,Salcedo:2002pr} it was
shown that a derivative expansion approach, suitably defined at finite
temperature, was appropriate to deal with this problem. We will use
this approach in the present work.

\subsection{Center symmetry in gluodynamics} 
\label{sec:center}

In pure gluodynamics at finite temperature one can use the center of
the gauge group to extend the periodic transformations to aperiodic
ones \cite{'tHooft:1979uj},
\begin{eqnarray}
g(\vec x, \frac1T ) =  z g (\vec x, 0 ) ,  \qquad z^{N_c} = 1  
\label{eq:ALGT1}
\end{eqnarray}
so that $z$ is an element of $\mathbb{Z}(N_c)$.  An example of such a
transformation (with $z=e^{i2\pi/N_c}$) in the Polyakov gauge is given
by
\begin{eqnarray}
g (x_4 ) =  \diag(e^{i  2 \pi x_4 n_j T / N_c  }),
\nonumber\\
n_1=1-N_c,\quad n_{j\ge 2}=1
\label{eq:ALGT}
\end{eqnarray}
and the gauge transformation on the $A_4$ component of the  gluon field is
\begin{equation}
A_4 \to A_4 +  \frac{2 \pi T }{ N_c}\diag(n_j)\,.
\end{equation}
Under these transformations both gluonic action, measure and boundary
conditions are invariant. The Polyakov loop, however, transforms as
the fundamental representation of the $\mathbb{Z}(N_c)$ group,
i.e. $\Omega \to z \Omega $, yielding $ \langle \Omega \rangle= z\langle \Omega \rangle $ and hence $\langle \Omega \rangle =0 $. More
generally, in the center symmetric or confining phase
\begin{eqnarray}
\langle \Omega^n \rangle =0 \quad \text{for} \quad n \neq k N_c\,,
\quad k\in\mathbb{Z} \,.
\label{eq:1.9}
\end{eqnarray} 
 Actually, this center symmetry is spontaneously broken {\em above} a
critical temperature, $T_D\approx 270\,$MeV for
$N_c=3$~\cite{Karsch:2001cy}.  The antiperiodic quark field boundary
conditions are not preserved under non trivial center transformations
since $q(\vec{x}, 1/T )\to g(\vec{x}, 1/T )q(\vec{x}, 1/T )=-z
g(\vec{x},0)q(\vec{x},0)$ instead of $-g(\vec{x},0)q(\vec{x},0)$.  A
direct consequence of such property is the vanishing of contributions
to the quark bilinear of the form\footnote{In this formula $\langle
\bar q ( n /T ) q( 0) \rangle$ denotes contributions to the quark
propagator including only paths which wind $n$ times around the
thermal cylinder. The average is for the quenched theory.}
\begin{eqnarray}
\langle \bar q ( n /T  ) q( 0) \rangle =0 \quad\text{for} \quad n
\neq kN_c, \quad k\in\mathbb{Z}
\label{eq:1.10}
\end{eqnarray} 
(in the confining phase) since under the large aperio\-dic
transformations given by Eq.~(\ref{eq:ALGT1}) $ \bar q ( n /T  ) q(
0) \to z^{-n} \bar q ( n /T  ) q( 0) $. This generates an exact
selection rule in quenched QCD.  The center symmetry is
explicitly broken by the presence of dynamical quarks and the choice
of an order parameter for confinement is not obvious
\cite{Fukushima:2002bk}. As a consequence the selection rule implied
by eq. (\ref{eq:1.10}) is no longer fulfilled. Nevertheless such
selection rule becomes relevant to chiral quark models in the large
$N_c$ limit and departures from it are found to be suppressed within
chiral quark models in the large $N_c$ limit at low temperatures, due
to the spontaneous breaking of chiral symmetry which generates heavier
constituent quarks from light current quarks.\footnote{We emphasize
that our use of the approximate rule is in contrast to the so-called
canonical ensemble description of QCD where, upon projection, triality
is assumed to be exact even in the presence of dynamical quarks. See
e.g. the discussion in \cite{Fukushima:2002bk}.} This issue will be
analyzed along this paper.

\subsection{Chiral quark models at finite temperature} 
\label{sec:cqmod}

Chiral quark models have been used in the past to provide some
semiquantitative understanding of hadronic features in the low energy
domain. At zero temperature chiral quark models are invariant under
{\em global} $\text{SU}(N_c)$ transformations. There has always been
the question of how the corresponding constituent quark fields
transform under {\em local} color transformations or whether a
physi\-cal gauge invariant definition can be attached to those
fields~\cite{Lavelle:1995ty}. If we assume that they transform in the same way as bare
quarks, it seems unavoidable to couple gluons to the model in the
standard way to maintain gauge invariance as done in previous works
(see e.g. Refs.~\cite{Espriu:1989ff,Bijnens:1992uz}). These gluon
effects are treated within perturbation theory at $T=0$. This
approximation induces some sub-leading corrections in the calculation
of color singlet states where the effects of confinement can be almost
completely ignored for the low lying states
\cite{Glozman:1995fu}. This perturbative gluon dressing also complies
with the interpretation that the whole quark model is defined at a low
renormalization scale, from which QCD perturbative evolution to high
energies processes can be successfully
applied~\cite{RuizArriola:2002wr}. When going to finite temperature,
chiral quark models predict already at the one loop level a chiral
phase transition~\cite{Bernard:1987ir,Christov:1991se} at realistic
temperatures. However, even at low temperatures single quark states
are excited what is obviously not very realistic for it means that the
hot environment is in fact a hot plasma of quarks. On the other hand
since the constituent quark mass is about a factor of 2 larger than
the pion mass, pion loops dominate at low
temperatures~\cite{Florkowski:1996wf} (for a review see
e.g. Ref.~\cite{Oertel:2000jp}), as expected from chiral perturbation
theory~\cite{Gasser:1986vb,Gerber:1988tt}. In the present work we will
deal with two chiral quark models, the Nambu--Jona-Lasinio (NJL)
model~\cite{Klevansky:1992qe,Christov:1995vm,Alkofer:1994ph} where
quarks are cha\-racterized by a constant constituent mass in the
propagator due to the spontaneous breaking of chiral symmetry and the
recently proposed spectral quark model
(SQM)~\cite{RuizArriola:2001rr,RuizArriola:2003bs,RuizArriola:2003wi,Megias:2004uj}
where the notion of analytic confinement is explicitly verified. For
completeness we review briefly the corresponding effective action
below.  One common and attractive feature of chiral quark models is
that there is a one-to-one relation to the large $N_c$ expansion and
the saddle point approximation of a given path integral both at zero
and at finite temperature.

\subsubsection{The NJL model} 

The NJL Lagrangian as will be used in this paper reads in Minkowski space\footnote{We use Bjorken-Drell convention throughout the paper.} 
\begin{eqnarray} 
{\cal L}_{\rm NJL} &=& \bar{q} (i\slashchar\partial 
- \hat{M}_0 )q \nonumber \\ &+& {G \over 2}\sum_{a=0}^{N_f^2-1}
\left( (\bar{q}\lambda_a q)^2 +(\bar{q}\lambda_a i \gamma_5 q)^2
\right)
\end{eqnarray} 
where $q=(u,d,s, \ldots )$ represents a quark spinor with $N_c $
colors and $N_f$ flavors.  The $\lambda$'s are the Gell-Mann flavor
matrices of the $U(N_f)$ group and $\hat M_0= \diag (m_u, m_d,
m_s,\ldots) $ stands for the current quark mass matrix.  In the
limiting case of vanishing current quark masses the classical
NJL-action is invariant under the global $U(N_f)_R \otimes U(N_f)_L $
group of transformations. Using the standard bosonization
procedure~\cite{Eguchi:1976iz} it is convenient to introduce auxiliary
bosonic fields $(S,P,V,A)$ so that after formally integrating out the
quarks one gets the effective action\footnote{Obviously at finite
temperature the quark fields satisfy antiperiodic boundary conditions
whereas the bosonized fields obey periodic boundary conditions.}
\begin{eqnarray}
\Gamma_{\rm NJL} [S,P] &=&-i N_c {\rm Tr} \log \left( i {\bf
D} \right)  \nonumber \\ &-& {1\over 4G } \int d^4 x   \,{\rm tr}_f \left( S^2
+ P^2 \right) \,.
\label{eq:eff_ac_njl} 
\end{eqnarray} 
We use ${\rm Tr}$ for the full functional trace, $\tr_f $ for the
trace in flavor space, and $\tr_c $ for the trace in color
space. Here, the Dirac operator is given by
\begin{eqnarray}
i {\bf D} &=& i\slashchar{\partial} - {\hat M_0} - \left( S + i
\gamma_5 P \right) \,.
\label{eq:dirac_op_njl} 
\end{eqnarray}
The divergencies in Eq.~(\ref{eq:eff_ac_njl}) from the Dirac
determinant can be regularized in a chiral gauge invariant manner by
means of the Pauli-Villars method, although the issue of
regularization is of little relevance at finite
temperature~\cite{Christov:1991se} for $T \ll \Lambda $. This model is
known not to confine and to produce a constituent quark mass $M \sim
300 {\rm MeV}$ due to the spontaneous breaking of chiral symmetry at
zero temperature. The Goldstone bosons can be parameterized by taking
\begin{eqnarray}
S+ iP = \sqrt{U} \Sigma \sqrt{U}   
\end{eqnarray} 
with $U $ a unitary matrix (see Eq.~(\ref{eq:pionU}) ) with $
\Sigma^\dagger = \Sigma $, and one can use that $\Sigma = M + \phi $
with $\phi$ the scalar field fluctuation. The partition function for this model can be written as 
\begin{eqnarray}
Z_{\rm NJL} = \int DU D\Sigma \, e^{i \Gamma_{\rm NJL} [ U , \Sigma ]} \,.
\label{eq:Z_njl} 
\end{eqnarray} 

By minimizing $\Gamma_{\rm NJL}$ one gets $S=M$, which generates the
spontaneous breaking of chiral symmetry, and one obtains the gap
equation
\begin{equation}
\frac{1}{G} = -i4N_c \sum_i c_i \int \frac{d^4 k}{(2\pi)^4} \frac{1}{k^2-M^2-\Lambda_i^2} \,,
\label{eq:gap_eq}
\end{equation}
where the Pauli-Villars regularization has been used. The
Pauli-Villars regulators fulfill $c_0=1$, $\Lambda_0=0$ and the
conditions $\sum_i c_i=0$, $\sum_i c_i \Lambda_i^2=0$, in order to
render finite the logarithmic and quadratic divergencies,
respectively. In practice it is common to take two cutoffs in the
coincidence limit $\Lambda_1 \to \Lambda_2 = \Lambda$ and hence
$\sum_i c_i f(\Lambda_i^2) = f(0)-f(\Lambda^2) + \Lambda^2 f^\prime
(\Lambda^2)$.

\subsubsection{The SQM model}

In the SQM  the effective action reads 
\begin{eqnarray}
\Gamma_{\rm SQM}[U] =- i N_c \int d \omega \rho(\omega) {\rm Tr} \log
\left( i {\bf D} \right),
\label{eq:eff_ac_sqm} 
\end{eqnarray} 
where the Dirac operator is given by 
\begin{eqnarray}
i {\bf D} &=& i\slashchar{\partial} - \omega U^{\gamma_5} - {\hat M_0} 
\label{eq:dirac_op_sqm} 
\end{eqnarray} 
and $\rho(\omega)$ is the spectral function of a generalized Lehmann
representation of the quark propagator with~$\omega$ the spectral mass
defined on a suitable contour of the complex
plane~\cite{RuizArriola:2001rr,RuizArriola:2003bs,RuizArriola:2003wi,
Megias:2004uj}. The use of certain spectral conditions guarantees
finiteness of the action. The matrix  $ U =
u^2 = e^{ { i} \sqrt{2} \Phi / f } $ ( $f$ is the pion weak decay
constant in the chiral limit) is the flavor matrix representing the
pseudoscalar octet of mesons in the non-linear representation,
\begin{eqnarray} 
	\Phi = \left( \matrix{ \frac{1}{\sqrt{2}} \pi^0 +
	\frac{1}{\sqrt{6}} \eta & \pi^+ & K^+  \cr  \pi^- & -
	\frac{1}{\sqrt{2}} \pi^0 + \frac{1}{\sqrt{6}} \eta & K^0  \cr 
	K^- & \bar{K}^0 & - \frac{2}{\sqrt{6}} \eta }
	\right) .
\label{eq:pionU}
\end{eqnarray}
A judicious choice of the spectral function based on vector meson
dominance generates a quark propagator with no-poles (analytic
confinement).  More details of the SQM at zero and finite
temperature relevant for the present paper are further developed at
Appendix~\ref{sec:sqm}. The partition function for the SQM can be written as 
\begin{eqnarray}
Z_{\rm SQM} = \int DU e^{i \Gamma_{\rm SQM} [ U ]} \,.
\label{eq:Z_sqm} 
\end{eqnarray}

\subsection{The Polyakov-Chiral Quark Model}
\label{sec:pcqm}

As we will show in Sect.~\ref{sec:problem} there is a conflict between
large gauge invariance at finite temperature, reviewed in the previous
Sects. \ref{sec:large-gauge} and \ref{sec:center}, and the standard
chiral quark models presented in Sect.~\ref{sec:cqmod}. The chiral
quark model coupled to the Polyakov loop that will be motivated in
Sect.~\ref{sec:polyakovcoupling} and analyzed in the rest of this
paper synthesizes the solution and corresponds to simply make the
replacement
 \begin{eqnarray}
\partial_4 \to \partial_4 - i A_4   
\end{eqnarray} 
in the Dirac operators, eq.~(\ref{eq:dirac_op_njl}) and
eq.~(\ref{eq:dirac_op_sqm}), and integrating further over the $A_4$
gluon field in a gauge invariant manner~\cite{Reinhardt:1996fs}
yielding a generic partition function of the form
\begin{eqnarray}
Z = \int DU D\Omega  \, e^{i \Gamma_G [\Omega]} e^{i \Gamma_Q [ U , \Omega ]} 
\label{eq:Z_pnjl} 
\end{eqnarray} 
where $DU$ is the Haar measure of the chiral flavor group $SU(N_f)_R
\times SU(N_f)_L $ and $D\Omega$ the Haar measure of the color group
$SU(N_c)$, $\Gamma_G $ is the effective gluon action whereas
$\Gamma_Q$ stands for the quark effective action. If the gluonic
measure is left out $A_4=0$ and $\Omega=1$ we recover the original
form of the corresponding chiral quark model, where there exists a
one-to-one mapping between the loop expansion and the large $N_c$
expansion both at zero and finite temperature. Equivalently one can
make a saddle point approximation and corrections thereof. In the
presence of the Polyakov loop such a correspondence does not hold, and
we will proceed by a quark loop expansion, i.e. a saddle point
approximation in the bosonic field $U$, keeping the integration on the
Polyakov loop $\Omega$. The work of Ref.~\cite{Fukushima:2003fw}
corresponds to make also a saddle point approximation in $\Omega$. In
Section~\ref{sec:oneloop} we stick to the one loop approximation and
keep the group integration.  This is the minimal way to comply with
center symmetry at low temperatures.  Although in principle
$\Omega(x)$ is a local variable, in what follows we will investigate
the consequences of a spatially constant Polyakov loop. In this case
the functional integration $D\Omega$ becomes a simple integration over
the gauge group $d\Omega$. The issue of locality is reconsidered in
Section \ref{sec:local}.

\section{Unnaturalness of chiral quark models at finite temperature} 
\label{sec:problem} 

In this section we analyze the problem of chiral quark models at
finite temperature, its interpretation in terms of thermal Boltzmann
factors as well as the corresponding conflicts with Chiral
Perturbation Theory at finite temperature.

\subsection{The problem}

As already mentioned, chiral quark models at finite temperature have a
pro\-blem since, even at low temperatures, excited states with any
number of quarks are involved, whether they can form a color singlet
or not. This is hardly a new observation, the surprising thing is that
nothing has been done about it so far, attributing the failure to
common diseases of the model, such as the lack of confinement. To
illustrate this point in some more detail we will use a constituent
quark model like the NJL model, where the quark propagator has a
constant mass. To be specific, let us consider as an example the
calculation of the quark condensate for a single flavor in
cons\-tituent quark models with mass $M$. At finite temperature in the
Matsubara formulation we have the standard rule
\begin{eqnarray}
\int \frac{d k_0}{2\pi}  
F(\vec k, k_0) \to i T
\sum_{n=-\infty}^\infty  F( \vec k,i \omega_n )
\end{eqnarray}
with $\omega_n$ the fermionic Matsubara frequencies, $ \omega_n = 2 \pi T (
n+1/2 ) $. For the discussion in this and forthcoming sections it is
convenient to elaborate this formula a bit further.  Using Poisson's
summation formula
\begin{eqnarray}
\sum_{m=-\infty}^\infty F ( m ) = \sum_{n=-\infty}^\infty %\left\{
\int_{-\infty}^\infty d x F ( x ) e^{ i 2\pi x n } %\right\} 
\label{eq:poisson}
\end{eqnarray} 
one gets the rule 
\begin{equation}
\int \frac{d k_0}{2\pi} F( \vec k, k_0 ) 
%\to i T \sum_{n=-\infty}^\infty  F(\vec k, i \omega_n ) = 
\to i \sum_{n=-\infty}^\infty (-1)^n \int \frac{d k_4}{2\pi} F( \vec
k, i k_4 ) e^{ i n k_4 / T } \,. \nonumber 
\label{eq:rule_pois}
\end{equation}
In terms of the Fourier transform, one obtains for a finite
temperature fermionic propagator starting and ending at the same point,
\begin{eqnarray}
\tilde{F}(x; x) \to  \sum_{n=-\infty}^\infty (-1)^n \tilde{F}( \vec
x,x_0+in /T ;\vec x,x_0) \,.
\label{eq:2.4}
\end{eqnarray}
Note that the zero temperature contribution corresponds to the term
$n=0$ in the sum. From a path integral point of view, the zero
temperature term comes from contractile closed paths whereas thermal
contributions come from closed paths which wind $n$ times around the
thermal cylinder. For fermions, each winding picks up a $-1$ factor.

For the (single flavor) condensate we get\footnote{In what follows we
use an asterisk as upperscript for finite temperature quantities,
i. e. ${\cal O}^* = {\cal O}_T $.}
\begin{eqnarray}
\langle \bar q q \rangle^* &=& 
-i N_c \sum_{n=-\infty}^\infty (-1)^n \tr_{\text{Dirac}} S (x)
\Big|_{x_0=i n /T } \nonumber \\ 
&=&  -i 4 M N_c
\sum_{n=-\infty}^\infty \int \frac{d^4 k}{(2\pi)^4} \frac{e^{ -i k
\cdot x } (-1)^{n} }{k^2 - M^2 } \Big|_{x_0=i n /T } \nonumber \\
&=& \langle \bar q q \rangle - 2 \frac{N_c M^2 T }{\pi^2} 
\sum_{n = 1}^\infty  \frac{(-1)^n}{n}   K_1 ( n M  /T ) \,.
\label{eq:cond}
\end{eqnarray} 
In writing the previous formula, finite cut-off corrections, appearing
in the chiral quark models such as the NJL model at finite temperature
have been neglected. This is not a bad approximation provided the
temperature is low enough $ T \ll \Lambda $ (typically one has $
\Lambda \approx 1\,$GeV so even for $ T \approx M \approx 300\,$MeV the
approximation works). At low temperatures we may use the asymptotic
form of the Bessel function \cite{Abramowitz:1970bk}
\begin{eqnarray}
K_n (z) \sim  e^{-z} \sqrt{\frac{\pi}{2z}} 
\label{eq:bess_asy} 
\end{eqnarray} 
to get for the leading contribution, 
\begin{eqnarray}
\langle \bar q q \rangle^* &\sim& \langle \bar q q \rangle + 4N_c 
\left(\frac{ M T }{2\pi} \right)^{3/2} e^{-M / T} \, . 
\end{eqnarray}
This means a rather flat dependence on temperature for $ T \lesssim M
$. (Numerically, the correction is about $ 1 \% $ for $ T \approx
100\,$MeV for $ M = 300\,$MeV and $ \langle \bar q q \rangle \approx -
(240 \,\text{MeV})^3 $). The strong attractive interaction which
causes chiral dynamical symmetry breaking is reduced at finite
temperature and the energy is decreased by a decreasing cons\-tituent
quark mass $M^*$, eventually leading to a chiral phase
transition~\cite{Bernard:1987ir,Christov:1991se}, the critical
temperature is $ T\approx 200\,$MeV.\footnote{The minimization can be
written as the equation $ \langle \bar q q \rangle^* (M^*) / M^*
=\langle \bar q q \rangle (M) / M $, so one has to know the mass
dependence of the condensate at zero temperature.} The coincidence of
this number with lattice simulations has been considered a big success
of chiral quark models and has triggered a lot of activity in the past
(see e.g. Ref.~\cite{Oertel:2000jp} and references therein). We show
below that this apparent success might be completely accidental, as it
does not incorporate basic physical requirements. 

\subsection{Interpretation} 

An interpretation of the previous formula for the condensate is in
terms of statistical Boltzmann factors. Using the definition of the
quark propagator in coordinate space
\begin{eqnarray}
S(x) = \int \frac{d^4 k}{(2\pi)^4}\frac{e^{-i k \cdot x}}{\slashchar{k} - M} = 
\left( i \slashchar{\partial} + M \right) \Delta ( x  ) 
\end{eqnarray}
with 
\begin{eqnarray}
\Delta(x ) = \int \frac{d^4 k}{(2\pi)^4}\frac{e^{-i k \cdot x}}{k^2 -
  M^2} = \frac{M^2}{4 \pi^2 i } \frac{K_1 ( \sqrt{-M^2 x^2}
  )}{\sqrt{-M^2 x^2}}\,,
\end{eqnarray}
at low temperature 
we get 
\begin{eqnarray}
S( \vec x ,i /T ) \sim e^{-M /T }  \,.
\end{eqnarray}
Thus, for $\langle{\bar q}q\rangle^*$ and up to prefactors, we have the
exponential suppression for a single quark propagator at low
temperature.  Using Eq.~(\ref{eq:cond}) and Eq.~(\ref{eq:bess_asy})
the quark condensate can be written in terms of Boltzmann factors with
a mass formula $ M_n = n M $ corresponding to any number of quark
states.

One might object against the previous interpretation by arguing that
these factors only reflect in the Euclidean coordinate space the pole
of the propagator in Minkowski momentum space, and hence that they are
a natural consequence of the lack of confinement. While the former
statement is true, in the sense that singularities in Minkowski
momentum space can be seen at large Euclidean coordinate values, the
conclusion drawn from there is incorrect. As shown in
Ref.~\cite{RuizArriola:2003wi} (see Appendix~\ref{sec:sqm}) quark
propagators with no poles but cuts can also produce a Boltzmann factor
{\it without} prefactors, as it should be.\footnote{Actually, the
previous counter example shows that the lack of confinement has more
to do with the presence of the exponential prefactors which are
related to the available phase space.} To the same level of
approximation, i.e. one quark loop, in the SQM we get (see
Appendix~\ref{sec:sqm} for details)
\begin{eqnarray}
\frac{\langle \bar q q \rangle^*}{\langle \bar q q \rangle} &=& \tanh
\left( M_S  /4T \right) \nonumber \\ &=& 
1 - 2 e^{- M_S /2T } + 2  e^{- M_S/T  } +  \cdots 
\label{eq:cond_SQM_nopol}
\end{eqnarray}  
where the ``Boltzmann'' constituent mass can be identified with half
the scalar meson mass $ M = M_S / 2$.\footnote{This relation together
with the large $N_c$ quark-hadron duality relation $M_S=M_V$ discussed
in Ref.~\cite{Megias:2004uj} yields $M= M_V/2 \sim 385 {\rm MeV}$, a
reasonable value.} These calculations illustrate our main point and
can be extended to any observables which are color singlets in the
zero temperature limit; quark model calculations at finite temperature
in the one loop approximation gene\-rate {\it all} possible quark
states,
\begin{eqnarray}
{\cal O}^* = {\cal O} + {\cal O}_q e^{-M/T} + {\cal O}_{qq} e^{-2
M/T} + \cdots
\end{eqnarray}
While there is no doubt that the leading term~${\cal O}_q$ has a
Boltzmann factor corresponding to a single quark state, the term with
mass $ 2 M $ could in principle be equally a $qq$ diquark state or a
$\bar q q $ meson state. The latter possi\-bility should be discarded,
however. At one loop a $\bar q q$ pair can only come from the quark
line going upwards and then downwards in imaginary time
propagation. Since such a path does not wind around the thermal
cylinder it is already counted in the zero temperature term. The $qq$
contribution, instead, corresponds to the single quark line looping
twice around the thermal cylinder and is a proper thermal
contribution. This is confirmed below.  These Boltzmann factors
control the whole physics and temperature effects are sizeable for $ T
\approx M $.

\subsection{Conflicts with ChPT} 

Our observation on the Boltzmann factor is rather puzz\-ling because
it seems hard to understand how is it possible to generate non singlet
states by just increasing the temperature. The reason has to do with
the fact that the condensate itself is not invariant under
$\mathbb{Z}(N_c)$ transformations at finite temperature. For the
example of the condensate we trivially obtain
\begin{eqnarray}
\langle \bar q q \rangle^* &=& \sum_{n=-\infty}^\infty (-1)^n \langle
\bar q(x_0 ) q (0) \rangle \Big|_{x_0=i n /T  }
\end{eqnarray} 
i.e., the condensate at finite temperature can be written as a
coherent sum of nonlocal quark condensates at zero temperature.  If we
make a gauge transformation of the central type, we get
\begin{eqnarray}
\langle \bar q q \rangle^* &\to & \sum_{n=-\infty}^\infty (-z)^n \langle
\bar q(x_0 ) q (0) \rangle \Big|_{x_0=i  n /T  }
\label{eq:2.14}
\end{eqnarray}
i.e., the condensate can be decomposed as a sum of irreducible
representations of a given triality $n$. Thus, the state with Boltzmann
factor $e^{-n M /T }$ is indeed a multiquark state. 

This avoids the paradox, and suggests that in order to make a
(centrally extended) gauge invariant definition of the condensate we
could simply discard from the sum those terms which do not have zero
triality, i.e. we would get
\begin{eqnarray}
\langle \bar q q \rangle^* \Big|_{\text{singlet}} &= &
\sum_{n=-\infty}^\infty (-1)^{nN_c} \langle \bar q(x_0 ) q (0) \rangle
\Big|_{x_0=i N_c n /T  } 
\label{eq:2.15}
\end{eqnarray}
This would generate as a first thermal correction a term with a Boltzmann factor corresponding to mass $ N_c M $
(a baryon) which is obviously very much suppressed. Since a quark loop
generates a dependence proportional to $N_c$ we would obtain a $ N_c
e^{-M N_c / T } $ dependence.

Another problem now comes from comparison with the expectations of
chiral perturbation theory at finite
temperature~\cite{Gasser:1986vb}. In the chiral limit, i.e., for $
m_\pi \ll 2 \pi T \ll 4 \pi f_\pi $ the leading thermal corrections to
the quark condensate for $N_f=2$, for instance, are given by
\begin{eqnarray}
\langle \bar q q \rangle^* \Big|_{\text{ChPT}} &= & \langle \bar q q
\rangle \left[ 1- \frac{T^2 } {8 f_\pi^2} - \frac{T^4 } {384 f_\pi^4}
+ \cdots \right]
\label{eq:2.16}
\end{eqnarray} 
Thus, the finite temperature correction is $N_c$-suppressed as
compared to the zero temperature value, since $f_\pi^2$ scales as
$N_c$. This feature remains for finite pion mass, and is generic to
any thermal correction in ChPT; the dominant contribution comes from
quantum pionic fluctuations and not from quark thermal excitations.
Although the previous formula predicts a lowering of the quark
condensate, it cannot describe the chiral phase transition since ChPT
assumes from the start a non vanishing chiral condensate. In this
sense, the scaling behavior of the critical temperature with $f_\pi$
and therefore with $N_c$ suggested from direct extrapolation of the
formula can only be regarded as an assumption.

At this point we should remind that the mechanism by which the chiral
symmetry is restored at finite temperature in standard chiral quark
models in the one quark loop approximation is quite different from the
trend deduced from ChPT based mainly on pion loops. While in the first
case it is due to populating states of the Dirac levels with the
Fermi-Dirac thermal factor and a sudden decrease of the constituent
quark mass gap $2M$, in ChPT the ``phase transition'' is merely due to
large quark-antiquark excitations with the lightest pion quantum
numbers with a fixed gap (otherwise ChPT method cannot be
applied). These two pictures of the chiral symmetry restoration are
not dual to each other; the $N_c$ behavior of the critical
temperature is different since in chiral quark models one has $T_c
\sim M \sim N_c^0 $ while in ChPT the extrapolated value of the
``critical temperature'' is $T_c \sim 2 \sqrt{2} f_\pi \sim
\sqrt{N_c}$. Quantum fluctuations have been included in chiral quark
models at finite temperature~~\cite{Florkowski:1996wf} (for a review
see e.g. Ref.~\cite{Oertel:2000jp}) and they are known to be $1/N_c $
suppressed. Actually, the sub-leading $1/N_c$ contribution reproduces
the first term of ChPT, eq.~(\ref{eq:2.16}), thus largely dominating
at low temperatures. Taking into account that ChPT by itself and more
refined approaches incorporating meson resonance
effects~\cite{Pelaez:1998vx,Pelaez:2002xf} provide a similar values of
the ``critical temperature'' quite close to the lattice
predictions~\cite{Karsch:1998hr} for dynamical fermions and
extrapolated to the chiral limit, one may wonder what is the meaning
of the mean field quark chiral phase transition predicted in the
past~\cite{Bernard:1987ir,Christov:1991se,Oertel:2000jp} and which has
become a justification for chiral quark models at finite temperatures.
These problems are also common to models where quarks and mesons are
regarded as independent degrees of freedom.

We will see in the rest of the paper how a convenient $N_c$
suppression of quark thermal corrections arises naturally when a color
source, the Polyakov loop, is coupled to the chiral quark model and
subsequent projection onto color neutral states is carried out. In
this scenario one would have a large transition temperature $T_c \sim
N_c M $ due to quarks, i.e. no symmetry restoration due to filling in
the states above the Dirac levels in the absence of dynamical gluons
and in the quenched approximation (Polyakov cooling). Non perturbative
gluonic corrections modify this picture; they do predict instead a
critical temperature roughly equal the deconfinement phase transition,
$T_c = T_D$. Finally, pion loops are protected from additional
suppressions, so that the final result will be fully compatible with
the ChPT behavior at low temperature.

\section{Coupling the Polyakov Loop in Chiral Quark Models}
\label{sec:polyakovcoupling} 

\subsection{General considerations}

As we have said, one can formally maintain gauge invariance at zero
temperature by coupling gluons to the model. In the spirit of the
model these degrees of freedom should be treated within perturbation
theory, since the constituent quarks carry some information on
non-perturbative gluon effects (see
e.g. Ref.~\cite{Espriu:1989ff,Bijnens:1992uz} for explicit
calculations in the low energy limit). At finite temperature the
situation is radically different; a perturbative treatment of the
$A_0$ component of gluon field manifestly breaks gauge invariance
(namely, under large gauge transformations).  The consequences of
treating such a coupling non-perturbatively in the case of a constant
$A_0$ field are straightforward and enlightening (see below for a
discussion on the $x$-dependent case).

Actually, in a more general context, the Polyakov loop appears
naturally in any finite temperature calculation in the presence of
minimally coupled vector fields within a derivative expansion or a
heat-kernel expansion approach. In this case, as shown in
\cite{Garcia-Recio:2000gt,Megias:2002vr}, a local one loop quantity,
such as the effective Lagrangian or an observable, takes the form
\begin{equation}
{\cal L}(x)=\sum_n \tr\left[ f_n(\Omega(x)){\cal O}_n(x)\right] \,,
\label{eq:3.1}
\end{equation}
where $\tr$ acts on all internal degrees of freedom, $n$ labels all
possible local gauge invariant operators ${\cal O}_n(x)$
(i.e. containing covariant derivatives), possibly with brea\-king of
Lorentz symmetry down to rotational symmetry, and $f_n(\Omega(x))$ are
temperature dependent functions of the Polyakov loop which replace the
numerical coefficients present in the zero temperature case. In this
general context $\Omega(x)$ would be the local Polyakov loop of all
mini\-mally coupled fields.\footnote{As noted below, in a model with
vector mesons, there would be a corresponding flavor Polyakov
loop. Such a contribution is expected to be much suppressed due to the
large physical mass of the mesons.} In particular, a chemical
potential would give a contribution $e^{\mu /T }$. Here we can see
the necessity of the presence of $\Omega$ in (\ref{eq:3.1}): being
$\mu$ a constant, it gives no contribution in the covariant derivative
and so in ${\cal O}_n(x)$, therefore the chemical potential can only
act through the presence of the Polyakov loop in the expression. This
consideration also illustrates the breaking of gauge invariance in a
perturbative treatment of $\Omega$: $e^{ \mu /T }$ depends
periodically on the chemical potential, with period $2\pi i T$, this
is a consequence of the coupling of $\mu$ to the integer quantized
particle (or rather charge) number. Such periodicity would be spoiled
in a perturbative treatment. Note that such periodicity is equivalent
to one-valuedness of the functions $f_n$ in (\ref{eq:3.1}).

\subsection{Coupling the Polyakov Loop}

Coming back to chiral quark models with gluonic Polyakov loops, in
fact, the analogy with the chemical potential has been invoked in a
recent proposal of
K. Fukushima~\cite{Fukushima:2003fw}~\footnote{After our work was
sent for publication  Refs.~\cite{Ratti:2005jh,Ghosh:2006qh} appeared, extending
the results of Fukushima.}, which suggests coupling chiral quark
mo\-dels to the Polyakov loop at finite temperature in this way. Our
own approach is similar, except that, as in (\ref{eq:3.1}), we
consider a {\em local} Polyakov loop $\Omega({\vec x})$ coupled to the
quarks. This is what comes out of explicit one loop calculations
within a derivative expansion approach at finite temperature
\cite{Megias:2002vr,Megias:2003ui,Megias:2004bj}.  In those
calculations there is a loop momenta integration at each given $x$,
and the Polyakov loop appears minimally coupled, i.e., through the
modified fermionic Matsubara frequencies,
\begin{eqnarray}
\om_n = 2 \pi T (n+1/2 + \hat \nu) \,,
\label{eq:3.2}
\end{eqnarray}
which are shifted by the logarithm of the Polyakov loop
\begin{eqnarray}
\Omega = e^{i 2 \pi \hat \nu}\,, 
\end{eqnarray}
i.e. $ \hat \nu(\vec x) = A_4(\vec x) /(2 \pi T) $. In our
considerations, this is the only place where explicit dependence on
color degrees of freedom appears, so it is useful to think of $\hat \nu
$ as the corresponding eigenvalues. The effect of such a shift
corresponds to change Eq.~(\ref{eq:2.4}) into
\begin{eqnarray}
\tilde{F}(x; x) \to \sum_{n=-\infty}^\infty (-\Omega(\vec x))^n
\tilde{F}( \vec x,x_0+in /T ;\vec x,x_0) \,.
\label{eq:2.4a}
\end{eqnarray}

\begin{figure}[tbc]
\begin{center}
\epsfig{figure=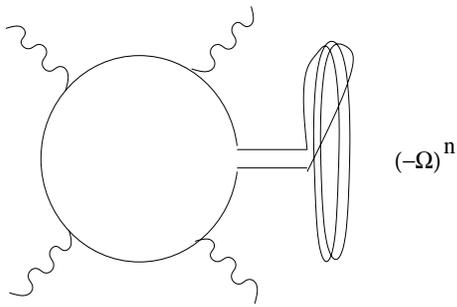,height=4cm,width=6cm}
\end{center}
\caption{Typical one quark loop diagram with a non trivial Wilson
line. For $n$ windings around the $U(1)$ compactified imaginary time
the quarks get a topological factor $\Omega^n $ in addition to the
Fermi-Dirac statistical factor $(-1)^n$. Wavy lines are external
fields. The total contribution to the diagram is obtained by summing
over all windings and tracing over color degrees of freedom.}
\label{fig:loop}
\end{figure}

The interpretation of this formula can be visualized in
Fig.~\ref{fig:loop}; in a one quark loop with any number of external
fields at finite temperature and with a non-trivial Polyakov line, the
quarks pick up a phase $(-1)$ due to Fermi-Dirac statistics, and a non
Abelian Aharonov-Bohm\footnote{This is an electric type of phase and
not the standard magnetic one. The name is nonetheless appropriate
since this electric phase was discussed first in the original AB
paper.} factor $\Omega$ each time the quarks wind around the
compactified Euclidean time. The total contribution to the diagram is
obtained by summing over all windings and tracing over color degrees
of freedom.

\subsection{Dynamical Polyakov loop}
\label{sec:dyn-pol}

The above prescription gives the contribution for a given gluon field
configuration, of which we have only retained the Polyakov
loop.\footnote{In addition, gluons appear also perturbatively through
the covariant derivative. This will produce perturbative gluon
exchange contributions as in the zero temperature case. We will not
consider those in this work.} The next step is to integrate the gluons
according to the QCD dynamics.  This implies an average over the local
Polyakov loop with some normalized weight
$\rho(\Omega;\vec{x}) d\Omega$. Here $d\Omega$ is the Haar measure of
SU($N_c$) and $\rho(\Omega;\vec{x})$ the (temperature dependent)
probability distribution of $\Omega(\vec{x})$ over the gauge group. The
emergence of the Haar measure of the integral representation of the
Yang-Mills partition function was explicitly shown in
Ref.~\cite{Reinhardt:1996fs}. Due to gauge invariance,
$\rho(\Omega)$ will be invariant under similari\-ty transformations,
and hence it is really a function of the eigenvalues of $\Omega$.  In
this section we will mainly remain within a quenched approximation and
so the weight follows from a pure Yang-Mills dynamics, in particular
the weight will be $\vec{x}$ independent, as we do not consider
external fields coupled to the gluons.\footnote{In
Sections~\ref{sec:local} and \ref{sec:unquenched} we will discuss some
implications about local corrections in the Polyakov loop and
unquenched results, respectively.}  In Yang-Mills dynamics (in four
dimensions and three colors) it its known to take place a first order
transition from a center symmetric phase to a broken symmetry or
deconfining phase. Note that this is a rather peculiar phase
transition where the symmetry is restored {\it below} the critical
temperature, just the opposite as the standard case. Since the
transition is discontinuous in observables such as the expectation
value of the Polyakov loop, the probability distribution
$\rho(\Omega)$ will also be discontinuous as a function of the
temperature at the critical temperature. In the confining phase
$\rho(\Omega)$ will be invariant under $\mathbb{Z}(N_c)$,
$\rho(\Omega)=\rho(z\Omega)$. In the deconfining phase, such
symmetry is spontaneously broken and one expects the Polyakov loop to
concentrate around one of the elements of the center, at random. The
coupling of dynamical quarks favors the perturbative value $\Omega=1$
($A_4=0$) as follows from computations of the effective potential at
high temperature \cite{Weiss:1980rj,Weiss:1981ev,Gross:1980br}. So in
that phase we expect to have $\Omega$ concentrated near
$\Omega=1$, which would be equivalent to no Polyakov loop in the
calculation.

Actually, one does not need the full distribution of $\Omega$ on
SU($N_c$), but only the marginal distribution of eigenvalues. Denoting
the Polyakov loop average by $\langle ~\rangle $, we have for a quark
observable
\begin{equation}
{\cal L}(x)=\sum_n \langle \tr_c f_n(\Omega) \rangle\, 
\tr\, {\cal O}_n(x) \,.
\label{eq:3.1a}
\end{equation}
Consistently with gauge invariance, the functions $f_n(\Omega)$ are just
ordinary functions $f_n(z)$ evaluated at $z=\Omega$ (e.g. $e^\Omega$)
hence, if $e^{i\phi_j}$, $j=1,\ldots,N_c$ are the eigenvalues of
$\Omega$
\begin{eqnarray}
\left\langle \frac{1}{N_c} \tr_c f(\Omega) \right\rangle &=&
\int_{\text{SU($N_c$)}}\!\!\! d\Omega \,\rho(\Omega)  \frac{1}{N_c}
\sum_{j=1}^{N_c}f(e^{i\phi_j})
\nonumber \\
&=& \int_{-\pi}^{\pi}\frac{d\phi}{2\pi}\widehat\rho(\phi) f(e^{i\phi})
\label{eq:one-body}
\end{eqnarray}
with
\begin{eqnarray}
\widehat\rho(\phi) &:=& 
\int_{\text{SU($N_c$)}}\!\!\! d\Omega \, \rho(\Omega) \frac{1}{N_c}
\sum_{j=1}^{N_c}2\pi\delta(\phi-\phi_j) \,.
\label{eq:4.7}
\end{eqnarray}
Equivalently, all that it is needed is the set of momenta of the
distribution, $\langle\tr_c(\Omega^n)\rangle$.

\subsection{Group averaging}

At sufficiently low temperature in the quenched theory we can go
further on the analytical side, since the distribution of the Polyakov
loop becomes just the Haar measure in this regime. As it will be
discussed in section~\ref{sec:gluon}, this fact is justified with
results based on strong coupling expansions and in one massive gluon
loop approximation. Actually, from eq. (\ref{eq:3.8}) we find that in
observables such as the quark condensate, the effect of
$\rho(\Omega)$ being different from unity is almost negligible for all
temperatures below the transition, implying that a Haar measure
distribution is an extremely good approximation in the confined
phase. We elaborate further on gluonic corrections in
section~\ref{sec:gluon}.

The corresponding density of eigenvalues of the SU($N_c$) group is
given by \cite{Miller:1972bk,Gross:1983pk}
\begin{eqnarray}
\frac{1}{N_c!}  
2\pi \delta\Big( \sum_{i=1}^{N_c} \phi_i \Big) \,
\prod_{i < j }^{N_c} | e^{i\phi_i} -  e^{i\phi_j} |^2
\prod_{i=1}^{N_c} \frac{d \phi_i}{2\pi}\,,
\label{eq:average1}
\end{eqnarray}
so $\widehat\rho(\phi)$ of (\ref{eq:4.7}) is simply
\begin{eqnarray}
\widehat\rho(\phi)=
  1 - \frac{2 (-1)^{N_c}}{N_c}  \cos( N_c \phi ) \,.
\end{eqnarray}
Using this result one can easily deduce the following useful formulas
for the average over the SU($N_c$) Haar measure
\begin{eqnarray}
\langle\tr_c(-\Omega)^n\rangle_{\text{SU($N_c$)}}  = \left\{\matrix{ 
N_c\,, & n=0 \label{eq:p1}\\
-1  \,, & n=\pm N_c \label{eq:p2} \\ 
  0 \,, & \text{otherwise} \label{eq:p3}\\
} \right.
\nonumber
\end{eqnarray}
When this is inserted in, e.g., Eq. (\ref{eq:3.1a}), one finds that
the effect is not only to remove the triality breaking terms, as in
Eq. (\ref{eq:2.15}), but additionally, the surviving thermal
contributions are $N_c$ suppressed as compared to the naive
expectation. This solves the second problem noted in Section
\ref{sec:problem}.

\subsection{Polyakov cooling mechanism}

In an insightful work, Fukushima \cite{Fukushima:2003fw} has modeled
the coupling of the Polyakov loop to chiral quarks, with emphasis in
the description of the deconfining and chiral phase transitions (or
rather, crossovers). The fact that the critical temperatures for both
transitions are nearly equal, according to lattice calculations
\cite{Karsch:2001cy}, finds a natu\-ral explanation in that model. This
follows from what we will call the {\em Polyakov cooling} mechanism,
namely, the observation that, upon introduction of coupling with the
Polyakov loop, any quark observable at temperature $T$ (below $T_D$)
roughly corresponds to the same observable in the theory without
Polyakov loop but at a lower temperature, of the order of $T/N_c$, as
already noted in \cite{Oleszczuk:1992yg}. This is a direct consequence
of triality conservation. As discussed for Eqs. (\ref{eq:2.14}) and
(\ref{eq:2.15}) at the end of Section \ref{sec:problem}, Boltzmann
weights $e^{-M/T}$ are suppressed in favor of $e^{-N_cM/T}$. An
extreme example of cooling would come from considering a U(1) gauge
theory in a confined phase in such a way that $\Omega$ is a completely
random phase coupled to the quark. This would be equivalent to a
uniform average of the variable $\hat\nu$ in Eq. (\ref{eq:3.2}) in the
interval $[0,1]$. Clearly, such an average completely removes the
discretization of the Matsubara frequencies and gives back the
continuum frequency of the zero temperature theory. The same extreme
cooling would obtain in a U($N_c$) gauge theory. In the SU($N_c$) case
the average is not so effective since the phases corresponding to each
of the $N_c$ colors are not changed independently, owing to the
restriction $\det\Omega=1$. The cooling mechanism will be substantially
modified in the unquenched theory, since sea quark loops allow to
create thermal (i.e., with $n$ different from zero in
e.g. Eq. (\ref{eq:2.4})) color singlet quark-antiquark pairs which
propagate without any direct influence of the Polyakov loop.

The way Polyakov cooling brings the chiral and deconfining critical
points to coincide is as follows. In the chiral theory without
Polyakov loop, the critical temperature of the chiral transition is
such that $T^{\Omega=1}_\chi<T_D$ yet $ N_c T_\chi^{\Omega=1}>T_D$.
Hence, in the theory with coupling to the Polyakov loop, one finds
that for $T<T_D$ Polyakov cooling acts, $\langle\bar{q}q\rangle^*$
becomes roughly that of $T/N_c$ which is below $T_\chi^{\Omega=1}$ and
chiral symmetry is broken. On the other hand, for $T>T_D$, Polyakov
cooling no longer acts and $\Omega$ quickly becomes unity, as in the
theory without Polyakov loop at the same temperature; since $T$ is
above $T_\chi^{\Omega=1}$, chiral symmetry is restored. As a
consequence the chiral transition is moved up and takes place at the
same temperature as the deconfining transition,
$T_\chi^{\langle\Omega\rangle}\approx T_D$. This result is consistent
with \cite{Coleman:1980mx} where it is shown that, at least in the
large $N_c$ limit, confinement implies chiral symmetry breaking.

We note a difference in our treatment of the Polyakov loop coupling
and that in \cite{Fukushima:2003fw}, namely, we use a local Polyakov
loop subject to thermal and quantum fluctuations, as described by the
distribution $\rho(\Omega;\vec{x}) d\Omega$. This is in contrast with
\cite{Fukushima:2003fw} where $\Omega$ is global and does not
fluctuate. Instead $\Omega$ is determined through a mean field
minimization procedure plus a specific choice of the allowed values
(orbit) of $\Omega$ on the group manifold. In this way a model is
obtained which is simple and at the same time can be used to address
key issues of QCD at finite temperature. Nevertheless let us argue why
such an approach needs to be improved. At sufficiently low temperature
the model in Ref.~\cite{Fukushima:2003fw} for the gluon dynamics
consist just of the invariant Haar measure on the gauge group,
therefore any group element is just as probable as any other. If one
takes some coordinates on the group manifold and makes a maximization
of the resulting probability density, one is actually maxi\-mazing the
Jacobian and the result will depend on the coordinates chosen. In the
deconfined phase the local Polyakov loop is still subject to
fluctuations (even in the thermodynamic limit). A different quantity
is $\overline\Omega$, the spatial average of the local
loop.\footnote{The quantity $\overline\Omega$ so defined does not lie
on the group manifold, so some prescription should be devised to map
it onto the group.} This is a global object by construction. Both
quantities, $\Omega(x)$ and $\overline\Omega$, have the same
expectation value, due to translational invariance, but
$\overline\Omega$ does not fluctuate in the thermodynamic limit. The
usual effective potential is so defined that its minimum gives the
correct expectation value, and so $\overline\Omega$, but it does not
give information on the fluctuations of $\Omega(x)$.

In the confining phase of the quenched theory triality is preserved,
hence, after gluon average, Eq. (\ref{eq:2.4a}) becomes
\begin{eqnarray}
&& \tilde{F}(x; x) \to
\label{eq:2.4b} \\
&&\sum_{n=-\infty}^\infty 
\langle (-\Omega(\vec x))^{nN_c} \rangle 
\tilde{F}( \vec x,x_0+inN_c /T ;\vec x,x_0) \,,
\nonumber
\end{eqnarray}
which is the quenching invoked in Section \ref{sec:problem}.

\section{One quark loop results} 
\label{sec:oneloop} 

The calculations outlined above in Sect.~\ref{sec:polyakovcoupling}
can be routinely applied to all observables. A more thorough and
systematic study will be presented elsewhere.  As an illustration we
show here low temperature results (i.e. retaining only the Haar
measure in the gluon averaging) for the quark condensate and the pion
weak and electromagnetic anomalous decays for their relevance in
chiral symmetry breaking, both for the NJL model as well as for the
SQM at the one quark loop level. In Section~\ref{sec:higher} we
discuss the structure of higher order corrections due to quark loops
while in Section~\ref{sec:gluon} dynamical gluonic effects are
considered. Corrections beyond the quenched approximation will be
explicitly computed in Section~\ref{sec:unquenched}. In
Ref.~\cite{Megias:2006prep} we compute the full chiral Lagrangian at
finite temperature at the one quark loop level.

\subsection{Results for Constituent Quark Models} 

To visualize the additional suppression we apply the previous result
to the calculation of the condensate at finite temperature. At the one
loop level we just make the substitution $ N_c(-1)^n \to \tr_c 
\langle (-\Omega)^n \rangle $. We get
\begin{eqnarray}
\langle \bar q q \rangle^* &=& -i 4 M \sum_{n=-\infty}^\infty 
\tr_c \langle(-\Omega)^n \rangle \int
\frac{d^4 k}{(2\pi)^4} \frac{e^{ -i k x} }{k^2- M^2 } \Big|_{x_0=i
n/T}  \nonumber \\ \label{eq:qq_wpl}  
\end{eqnarray} 
This yields 
\begin{eqnarray}
\langle \bar q q \rangle^* = \langle \bar q q \rangle + \frac{2 M^2 T
}{ \pi^2 N_c} K_1 ( N_c M/T ) + \cdots 
\end{eqnarray}
The dots indicate higher gluonic or sea quark effects.
Because $T$ is small we have further
\begin{eqnarray}
\langle \bar q q \rangle^* & \sim & \langle \bar q q \rangle 
+ 4 \left(\frac{ M T }{2\pi N_c} \right)^{3/2} e^{-N_c M / T} \, . 
\label{eq:cond_CQM}
\end{eqnarray} 
When compared to the ChPT result Eq.~(\ref{eq:2.16}) we see that the
$N_c$ suppression of the constituent quark loop model is consistent
with the expectations.

For the pion weak decay constant we obtain 
\begin{eqnarray}
f_{\pi}^*{}^2 &=& -i 4 M^2 \nonumber \\ &\times& 
\sum_{n=-\infty}^\infty \tr_c \langle(-\Omega)^n \rangle
\int \frac{d^4 k}{(2\pi)^4} \frac{e^{ -i k
\cdot x}  }{[k^2- M^2 ]^2} \Big|_{x_0 = i n /T}
 \nonumber \\ 
\end{eqnarray}
yielding 
\begin{eqnarray}
\frac{f^*_\pi{}^2}{f_\pi^2} &=& 1  - \frac{M^2 }{ \pi^2 f_\pi^2}
K_0 (N_c M / T) + \cdots   
\end{eqnarray} 

The $\pi^0 \to \gamma \gamma $ amplitude is given by
\begin{eqnarray}
F_{\pi \gamma \gamma }^* &=& i \frac{8 M^2 }{N_c f_\pi} \nonumber \\ &\times& 
\sum_{n=-\infty}^\infty 
\tr_c \langle(-\Omega)^n \rangle
\int \frac{d^4 k}{(2\pi)^4} \frac{e^{ -i k
\cdot x}  }{[k^2- M^2 ]^3} \Big|_{x_0 = i n /T} \,.
 \nonumber \\ 
\end{eqnarray}
Using the value obtained at zero temperature, $ F_{\pi \gamma \gamma}
= 1/ 4\pi^2 f_\pi $, consistent with the anomaly, we get
\begin{eqnarray}
\frac{F_{\pi \gamma \gamma }^*}{F_{\pi \gamma \gamma }} &=& 1-\frac{2M}{T}
K_1 ( N_c M / T) + \cdots
\end{eqnarray}
This obviously complies again to the fact that the leading low
temperature corrections should be encoded in pionic thermal
excitations rather than quark excitations. 

\subsection{Spectral Quark Model} 

In the spectral quark model one averages with a given spectral
function the previous result (\ref{eq:cond_SQM_nopol}) and including
the Polyakov loop average we get (see Appendix~\ref{sec:sqm} for details)
\begin{equation}
\frac{\langle \bar q q \rangle^*}{\langle \bar q q \rangle} = 1-
\frac{2}{N_c} e^{-N_c M_S / 2T } + \cdots
\label{eq:cond_SQM}
\end{equation}
For the pion weak decay constant we obtain
\begin{eqnarray}
\frac{f^*_\pi{}^2}{f_\pi^2} &=& 1-
\frac{1}{N_c}\left(2+\frac{N_c M_V}{T}\right) e^{-N_c M_V/ 2T } 
+ \cdots
\nonumber \\
\label{eq:f_SQM}
\end{eqnarray}
and the $\pi^0 \to \gamma \gamma $ amplitude is given by
\begin{widetext}
\begin{eqnarray}
\frac{F_{\pi \gamma \gamma }^*}{F_{\pi \gamma \gamma }} 
= 1- \frac{1}{6N_c}  
\left[ 12+\frac{ 6 N_c M_V}{T}
+\left(\frac{N_c M_V}{T}\right)^2\right] e^{-N_c M_V/2T}+ \cdots \nonumber \\ 
\label{eq:F_SQM}
\end{eqnarray}
\end{widetext}

\begin{figure*}[tbc]
\begin{center}
\epsfig{figure=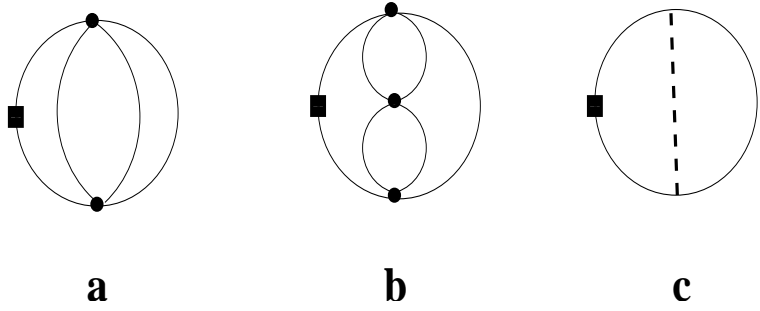,height=4cm,width=10cm}
\end{center}
\caption{Typical higher quark loop diagram for the quark condensate
operator $ \bar q q $. Quark lines with independent momenta may wind
n-times around the compactified Euclidean time, yielding a
Fermi-Polyakov factor $ (-\Omega)^n $. Triality conservation allows
the internal quark-antiquark lines to wind with opposite signs only
once, yielding an exponential suppression $ e^{- 2 M /T }$ for
diagram a). A similar suppression occurs for diagram b) if the
quark-antiquark windings happen at any of the bubbles. Diagram c)
corresponds to summing up all intermediate states with the same
quantum numbers and can be interpreted as a meson line.}
\label{fig:3loop}
\end{figure*}

\section{Corrections beyond one quark loop}
\label{sec:corrections}

In the previous sections we have restricted to the one quark loop
approximation for observables. This corresponds to the quenched
approximation within the model, and to some extent this provides an
oversimplified picture. In the present section we discuss the kind of
corrections that we expect to this approximation.

\subsection{Higher Quark Loop Corrections}
\label{sec:higher}

Going beyond the one quark loop approximation may require tedious
calculations (see e.g. Refs.~\cite{Florkowski:1996wf,Oertel:2000jp}
for explicit calculations in the standard NJL with no Polyakov
loop). However, some general features based on $N_c$ counting rules at
finite temperature can be deduced as follows. Let us, for
instance, consider the three loop diagram of Fig.~(\ref{fig:3loop}a)
contribution to the quark condensate in the NJL model in terms of
quark propagators.  Writing out for simplicity the Matsubara
frequencies only we have
\begin{eqnarray} 
{\rm Fig}.(\ref{fig:3loop} a) &=& 
\sum_{w^{(1)}, w^{(2)}) , w^{(3)}} S ( w^{(1)} ) \otimes S(
w^{(1)})  \\ &\otimes&  S( w^{(2)}) \otimes S( w^{(3)}) \otimes
S( w^{(1)}+ w^{(3)}- w^{(2)}) \nonumber 
\end{eqnarray} 
where $\otimes$ means tensor product in the Dirac and internal space
sense.  Using the Poisson's summation formula, Eq.~(\ref{eq:poisson}),
and going to Euclidean time space we get
\begin{widetext} 
\begin{eqnarray} 
{\rm Fig}.(\ref{fig:3loop} a) &=& 
\sum_{n_1,n_2,n_3} \langle \Omega^{n_1+n_2+n_3} \rangle \nonumber \\ &\times& 
 \int_{-\infty}^\infty d \tau_1 d \tau_3  S (
\tau_1 ) \otimes S( -\tau_1 - \tau_3 + n_1 /T + n_3 /T ) \otimes S( -\tau_3 + n_2 /T +n_3 /T ) \otimes S(
\tau_3 ) \otimes S( \tau_3 - n_3 /T ) \nonumber \\ &\sim &
\sum_{n_1,n_2,n_3} \langle \Omega^{n_1+n_2+n_3} \rangle e^{- M /T 
(|n_1|+|n_2|+|n_3|)}
\end{eqnarray} 
\end{widetext} 
For this diagram triality conservation implies, $ n_1+n_2+n_3 = k N_c
$ and the minimum argument of the exponent corresponds to take
$n_1=n_2=n_3= 0$, which is the zero temperature contribution. The next
thermal correction at low temperature is given by $n_1=0$ , $
n_2=-n_3= 1$ so the 3 loop diagram of Fig.~(\ref{fig:3loop}a) is
suppressed by a thermal factor $ e^{-2 M /T  } $, to be compared to
the one quark loop suppression $ e^{-N_c M /T } $. A similar thermal
suppression is obtained by inserting the standard bubble summation
which can be coupled to meson quantum numbers transforming the
argument of the exponent $ 2 M \to M_{\bar q q} $. Obviously, this
contribution becomes most important for the lightest pion state.
Actually, the quark-meson diagram in Fig.~(\ref{fig:3loop}b) looks as
a two loop bosonized diagram as shown in Fig.~(\ref{fig:3loop}c). For
such a bosonized diagram the previous argument becomes actually much
simpler, since the number of loops equals the number quark
propagators. The pion polarization operator, proportional to the pion
propagator can then be taken at zero temperature, since the most
important suppression comes from the quark lines not coupled to pion
quantum numbers.

For a bosonized diagram with L quark loops we have to consider L-fold
Matsubara generalization of the previous one quark loop correction
Eq.~(\ref{eq:2.4a}). Actually, the analysis becomes simpler in
coordinate space.  Regardless of the total number of quark propagators
we may choose to apply the Poisson's summation to $L$ quark
propagators. This can be seen by just using the formula
\begin{eqnarray}
\sum_{n,m=-\infty}^\infty \int_0^{1/T} d x_4 F( x_4 + n /T  
m /T  ) \nonumber \\ = \sum_{n=-\infty}^\infty \int_{-\infty}^\infty
d x_4 F( x_4 + n /T  )
\end{eqnarray}  
and its multidimensional generalization both in the sum and in the
integral sense. This effectively means that it is possible to remove
as many Poisson summations as coordinate integrals appear in the
expression. Using $ L= I-(V-1)$ and $ 4 V = E + 2I$ we also have
\begin{equation}
{\prod}_{i=0}^L \int d^4 z_i G^{2L} \sum_{n_1, \ldots , n_L}
{\prod}_{i=1}^L (-\Omega)^{n_i} S ( \vec x_i , t_i + i n_i /T 
)  \,. 
\end{equation} 
Actually, this rule does not depend on the precise form of the quark
interaction. At low temperatures, each quark line with an independent
Poisson index generates a constituent quark mass suppression. Thus,
the contribution to an observable can schematically be decomposed as
follows
\begin{equation} 
{\cal O}^* = \sum_L \sum_{n_1 , \ldots , n_L } {\cal O}_{n_1 \ldots n_L
} \langle \Omega^{n_1+ \cdots + n_L } \rangle e^{- ( |n_1| +
\cdots + |n_L| ) M /T } \,.
\label{eq:loop}
\end{equation} 
Triality conservation of the measure $ \Omega \to z \Omega $ at this
level implies
\begin{eqnarray} 
n_1 + \cdots + n_L = N_c k 
\label{eq:n-ality}
\end{eqnarray} 
with $ k$ an integer.  The dominant term in the previous expansion
is the one for which $ n_1 = \ldots = n_L =0$ with any arbitrary number
of quark loops $L$ and corresponds to the zero temperature
contribution. One also sees that for $L=1$ we only have contributions
from $n_1=k N_c$, which give the correction $e^{-N_c M /T  } $,
hence reproducing the results of Sect.~\ref{sec:oneloop}.  According
to Eq.~(\ref{eq:loop}), we can organize the thermal expansion at
finite but low temperatures. The most important contributions comes
from minimizing $ \sum_{i=1}^L | n_i | $ subjected to the triality
constraint, Eq.~(\ref{eq:n-ality}). At finite $T$ and for $N_c \ge 3 $
we have the leading temperature dependent contribution is given by $L
\ge 2$ and $ n_1 = -n_2 = 1 $ with $n_3= \cdots = n_L =0 $, which gives
a factor $ e^{-2 M /T } $ and corresponds to a $\bar q q $ singlet
meson state. This contribution has an additional $1/N_c$ power
suppression, as compared to the zero temperature contribution. For
$N_c=3$ the next term in the expansion would correspond to $L \ge 3$
and $ n_1= n_2 =n_3 =1 $ and yields a finite temperature suppression $
e^{-N_c M /T } $. For $N_c \ge 5 $ we would instead get $L \ge 4 $
and $ n_1=-n_2=n_3=n_4 = 1 $ and $ n_5 = \cdots n_L =0 $. Assuming $ N_c = 3$ 
we have\footnote{In the case without Polyakov loop one would have $ Z_{
q^{N_q} (\bar q q)^{N_m} } \sim \frac1{N_c^{N_m}} e^{-(2 N_m +N_q )
M / T} $ instead. So the leading contributions are those corresponding
to one quark state.}
\begin{eqnarray} 
Z_{\bar q q } &\sim&   \frac1{N_c} e^{-2 M / T} \\ 
Z_{qqq} & \sim&  e^{-N_c M / T} \\
Z_{qqq \bar q q } &\sim&   \frac1{N_c} e^{-(2+N_c) M / T} \\ 
&& \dots \\ 
Z_{ (\bar q q)^{N_M}  (qqq)^{N_B} } &\sim& \frac1{N_c^{N_M}} e^{-(2 N_M
+N_B N_c) M / T}
\end{eqnarray} 
Obviously, for $N_c =3$ the meson loop contribution dominates over the
baryon loop contribution.  The previous argument ignores completely
the quark binding effects so we should actually consider the relevant
meson mass $m$, thus in summary one would get 
 \begin{eqnarray} 
{\cal O} = 1 + \sum_{m} {\cal O}_{m} \frac1{N_c}e^{-m/T}  
+ \sum_{B} {\cal O}_B e^{-M_B/T } + \cdots \nonumber \\ 
\end{eqnarray} 
This is how quark-hadron duality works at finite temperature in chiral
quark models. As we see contributions of pion loops are the most
important ones, even though they are $1/N_c$ suppressed. Higher meson
states contribute next to the total observable at finite $T$. This is
what one naively expects and it is rewarding to see that such a
feature arises as a consequence of including the Polyakov loop into
the chiral quark model and subsequent projecting onto the gauge
invariant color singlet sector. 

Thus, at finite temperature there are the standard power like $1/N_c
\,e^{-2 M/T} $ suppression for meson loops accompanied by an exponential
suppression and a finite temperature exponential $ e^{-N_c M/T} $ for
baryon loops. Obviously the most important contributions at large
$N_c$ or low $T$ are those due to meson loops. We conclude from this
discussion that thermal pion loops are protected.

The previous discussion has concentrated on quark observables. For an observable like the Polyakov loop one would have instead
\begin{equation} 
\sum_L \sum_{n_1 , \ldots , n_L } {\cal O}_{n_1 \ldots n_L
} \langle \Omega^{1+n_1+ \cdots + n_L } \rangle e^{-  ( |n_1| +
\cdots + |n_L| ) M /T }
\label{eq:loop1}
\end{equation}
and
\begin{eqnarray} 
1+n_1 + \cdots + n_L = N_c k 
\label{eq:n-ality1}
\end{eqnarray}
The leading low temperature contribution (in this case there is no
zero temperature term) is then of the type $n_1=-1$,
$n_2=\cdots=n_L=0$, corresponding to a single antiquark loop screening
the charge of the test Polyakov loop. The leading term scales as
$e^{-M /T }$ and is controlled by the constituent quark mass. Unlike
the quark condensate case this behavior should remain unchanged by
pionic loops.

\subsection{Gluonic Corrections}
\label{sec:gluon}

Up to now we have chosen to represent the full dynamical gluonic
measure by a simple group integration. Unfortunately, we do not know
at present any general argument supporting the idea that there is a
low temperature exponential suppression of gluon degrees of freedom,
leaving only the Haar measure as the only remnant of gluon
dynamics. However, results based on strong coupling
expansions~\cite{Polonyi:1982wz,Gross:1983pk} and in one massive gluon
loop approximation~\cite{Meisinger:2001cq,Meisinger:2003id} do provide
such a suppression and indeed recent lattice findings confirm a
striking universality in all group representations and favoring the
simple group averaging dominance mechanism in gluodynamics below the
phase transition \cite{Dumitru:2003hp}. More specifically, one finds
both from lattice calculations \cite{Dumitru:2003hp} and from the
group measure that
\begin{equation}
\langle\widehat {\rm tr}_c \,
\widehat \Omega\rangle = 0
\end{equation}
in the confining phase for the Polyakov loop in the adjoint
representation. (In the group integration case, the previous formula
follows from (\ref{eq:7.4}) below). We stress that this result is not a
consequence of triality preservation since $\widehat \Omega$ is
invariant under 't Hooft transformations. The previous equation is
equivalent to $\langle |\mathrm{tr_c} \Omega|^2 \rangle = 1 $.  We
note in passing that in the mean field approximation
\cite{Fukushima:2003fw} $\langle |\mathrm{tr_c}\Omega|^2 \rangle$
vanishes instead, due to the absence of fluctuations.

We analyze now the two above mentioned models.

\subsubsection{Strong coupling expansion} 
\label{sec:strong-coupling}

The gluon potential at the leading order result of the strong coupling
expansion, for $N_c=3$, is taken as~\cite{Polonyi:1982wz,Gross:1983pk}
\begin{equation}
 -i\Gamma_G[\Omega] = 
V_{\text{glue}}[ \Omega ]\cdot a^3/T=-2(d-1)\,\mathrm{e}^{-\sigma a/T}
  \bigl|\mathrm{tr_c} \Omega \bigr|^2
\label{eq:G_potential_sce}
\end{equation}
with the string tension $\sigma=(425\,\text{MeV})^2$. At the mean
field level $V_{\text{glue}}$ leads to a first order phase
transition with the critical coupling $2(d-1)\mathrm{e}^{-\sigma
a/T_D}=0.5153$. One can fix the deconfinement transition temperature
as the empirical value $T_D=270\,\text{MeV}$ by choosing
$a^{-1}=272\,\text{MeV}$ \cite{Fukushima:2003fw}. The corresponding
mass is $m_G = \sigma a = 664\,\text{MeV}$.  At low temperatures we may
expand the exponential in powers of the gluon action,
\begin{equation}
 e^{i\Gamma_G} = 1 + i \Gamma_G - \frac{1}{2} \Gamma_G^2 + \cdots 
\end{equation}
which introduces an exponential suppression for $ e^{- m_G /T }$. For
a treatment based on an average over the Polyakov loop, the normalized
weight $\rho(\Omega) d\Omega$ suggested by the strong coupling
expansion will be
\begin{equation}
\rho(\Omega) = N\exp\left( 2(d-1)\, e^{-m_G/T}|\tr_c\Omega|^2\right)\,,
\label{eq:3.8}
\end{equation}
where $N$ is the normalization constant. Such distribution preserves
exact triality. At low temperature $\rho(\Omega)$ is close to unity
and the distribution coincides with the Haar measure, hence $\Omega$
is completely random with equal pro\-bability to take any group
value. At higher temperature $\rho(\Omega)$ tends to favor
concentration of $\Omega$ near the central elements of the group, with
equal probability.

This provides the following mass formula for the Boltzmann argument of the
exponential (in the notation of subsection \ref{sec:higher})
\begin{eqnarray}
{\cal M} = n N_c M_q + m M_{\bar q q } + l m_G 
\end{eqnarray} 
which clearly shows that the leading thermal contribution at low
temperatures is, again, provided by pion thermal loops, corresponding
to $n=l=0$ and $m=1$ due to $N_c M_q \gg m_G \gg M_{\bar q q
}=m_\pi$. Note that numerically, even the two pion contribution would
be more important than gluonic corrections.

\subsubsection{One massive gluon loop approximation} 

In a series of recent works~\cite{Meisinger:2001cq,Meisinger:2003id}
the equation of state has been deduced for a gas of massive gluons
with a temperature dependent mass in the presence of the Polyakov
loop, reproducing the lattice data quite accurately above the
deconfinement phase transition. The vacuum energy density reads
\begin{equation}
 V_{\text{glue}}[ \Omega ]= T \int \frac{d^3 k}{(2\pi)^3} \widehat{\rm
 tr}_{c} \log \left[ 1 - e^{-\omega_k /T } \widehat \Omega \right]
\end{equation}
where $ \omega_k = \sqrt{k^2 + m_G^2 } $, with $ m_G $ the gluons mass
and $\widehat \Omega $ and $\widehat {\rm tr}_c $ are the Polyakov
loop and the color trace in the adjoint representation respectively.
This expression was discussed with a temperature dependent mass in the
deconfined phase given by the plugging the Debye screening mass
$m_G(T)= T g(T) \sqrt{2} $, which at the phase transition, $T=T_c$,
takes the value $ m_G(T_c) = 1.2-1.3 T_c $. It is worth noticing that,
if one assumes a constant value for the gluon mass below the phase
transition one gets at low temperatures
\begin{equation}
 V_{\text{glue}}[ \Omega ]=- T \sum_{n=1}^\infty \frac1{n} (
\bigl|\mathrm{tr_c} \Omega^n \bigr|^{2} -1) \int \frac{d^3 k}{(2\pi)^3}
e^{-n \omega_k /T }
\end{equation}
where the identity
\begin{eqnarray}
\widehat {\rm tr}_c \, \widehat \Omega^n = \Bigl|\mathrm{tr}_c \Omega^n
\bigr|^{2}-1 
\label{eq:pol-adjoint}
\end{eqnarray}
has been used. Using the asymptotic representation of the Bessel
functions we see that, up to prefactors, a similar suppression of the
sort described in the strong coupling limit,
Sect.~\ref{sec:strong-coupling}, takes place.

\subsection{Local corrections in the Polyakov loop} 
\label{sec:local} 

Up to now we have assumed a constant $\Omega$ field in space in our
calculations. Quite generally, however, the Polyakov loop depends both
on the Euclidean time and the space coordinate, as it comes out of
explicit one loop calculations within a derivative expansion approach
at finite temperature
\cite{Megias:2002vr,Megias:2003ui,Megias:2004bj}.  In the Polyakov
gauge the temporal dependence becomes simple, but there is still an
unknown space coordinate dependence. In such a case, the previous
rules have to be modified, since Polyakov loop insertions carry finite
momentum, and the result depends on the ordering of these
insertions. If we still assume that the Polyakov loop is the only
color source in the problem, we are naturally lead to consider
Polyakov loop correlation functions. 
In the confining phase we expect
a cluster decomposition property to hold for any pair of
variables. 
A convenient model to account for Polyakov loop correlations is
\begin{eqnarray}
\langle \tr_c \Omega (\vec{x}) \; \tr_c
\Omega^{-1} (\vec{y}) \rangle = e^{-\sigma |\vec{x}-\vec{y}| /T} \,, 
\label{eq:corr-func}
\end{eqnarray} 
with $\sigma$ the string tension. This includes the correct screening
of the color charge at large distances due to confinement and is
consistent with (\ref{eq:7.5}) for two Polyakov loops at the same
point.
Thus, very different values of the spatial coordinate are suppressed,
and it makes sense to make a sort of local approximation within the
correlation length, and expand correlation functions in gradients in
that limited region of space.  Effectively, this corresponds to
replace the  volume to given confinement domain, by
means of the rule
\begin{equation}
\frac{V}{T} = \frac{1}{T} \int d^3 x \to   \frac{1}{T} \int d^3 x \, e^{-\sigma r /T}   = \frac{8 \pi T^2}{\sigma^3}  \,.
\label{eq:rule-vol}
\end{equation} 
In Ref.~\cite{Megias:2006prep} we will see explicitly that when
computing the low energy chiral Lagrangian by expanding the effective
action in derivatives of the meson fields there appear also gradients
of the Polyakov loop.  Actually, since we couple the coordinate
dependent Polyakov loop effectively as a $x$-dependent color chemical
potential our approach resembles a non-abelian generalization of the
local density approximation of many body physics in nuclear physics
and condensed matter systems, very much in the spirit of a density
functional theory.

\section{Results beyond the quenched approximation at low temperatures}
\label{sec:unquenched}

\subsection{General remarks}

The full Polyakov-Chiral quark model is given in Sect.~\ref{sec:pcqm}
by Eq.~(\ref{eq:Z_pnjl}). Therefore any expectation value is defined as
\begin{eqnarray}
\langle {\cal O}\rangle^* = 
\frac{1}{Z}\int DU D\Omega  
\, e^{i \Gamma_G [\Omega]} e^{i \Gamma_Q [ U , \Omega ]} 
\, {\cal O} 
\label{eq:O_pnjl} 
\end{eqnarray} 
with $\Gamma_G [\Omega]$ given in (\ref{eq:G_potential_sce}) and
$\Gamma_Q[U,\Omega]$ the quark contribution to the full action,
given by (\ref{eq:eff_ac_njl}) in the NJL model and
(\ref{eq:eff_ac_sqm}) for the SQM case. In the latter model the full
quark contribution coincides with the fermion determinant, while in
the NJL model there is an additional term arising from the
bosonization procedure,
\begin{equation}
e^{i\Gamma_Q[U,\Omega]} = \Det(i{\bf D})_\Omega \exp \left(
-\frac{i}{4G}\int d^4 x \, \tr_f (M-\hat{M}_0)^2\right) \,.
\label{eq:det_b_Q}
\end{equation}
(Note that here we have included in ${\bf D}$ the color degrees of
freedom.) 

In this section we gather all our results to provide an estimate of
the Polyakov loop expectation value at low temperatures as well as the
quark condensate. This is particularly interesting since in the
quenched approximation $\langle \tr_c \Omega \rangle =0$, due to
triality conservation. The fermion determinant does not conserve
triality, but we show below that at low temperatures the violation is
exponentially suppressed, so that it is still a good quantum number,
and the Polyakov loop can be used as an order parameter for center
symmetry in the same way as the chiral condensate provides a measure
of chiral symmetry restoration away from the chiral limit.

In order to go beyond the quenched approximation, we will evaluate the
fermion determinant in the presence of a slowly varying Polyakov loop
following the techniques developed in our previous
work~\cite{Megias:2002vr}. According to our discussion of
Sect.~\ref{sec:local} of local corrections, such an approximation
makes sense in a confining region where there are very strong
correlations between Polyakov loops. In the presence of the Polyakov
loop the quark contribution can be generally written as
\begin{equation}
e^{i\Gamma_Q[U,\Omega]} = e^{i\int d^4 x  {\cal L} (x, \Omega) } 
\end{equation} 
where ${\cal L}$ is the chiral Lagrangian as a function of the
Polyakov loop which will be computed at finite temperature in
Ref.~\cite{Megias:2006prep} in chiral quark models for non-vanishing
meson fields. For our purposes here only the vacuum contribution with
vanishing meson fields will be needed. 

\subsection{SQM model}

In our case it is simpler to consider first the SQM. We have
\begin{eqnarray}
e^{i\Gamma_Q[U,\Omega]}=\Det (i{\bf D})_\Omega = e^{ V B^*/T }     \,,
\end{eqnarray} 
where $V$ is the three dimensional volume and $-B^*$ the vacuum energy
density at finite temperature in the presence of the Polyakov
loop. The result for $B^*$ is quite simple and is listed in
(\ref{eq:A.19}) in Appendix~\ref{sec:sqm}.  At low temperatures we may
expand to get,
\begin{equation}
e^{ V B^*/T } = e^{ V B/T } \left[ 1 - \frac{V B}{T} e^{-M/T} \frac1{N_c} \tr_c
\left( \Omega + \Omega^{-1} \right)+ \cdots \right]
\end{equation}  
with $M=M_V/2$ the constituent quark mass in the SQM and $-B$ is the
vacuum energy density at zero temperature, $B= M_V^4 N_c N_f /192
\pi^2 = (0.2 {\rm GeV})^4 $ for three flavors (see Appendix
\ref{sec:sqm}). The calculation of observables requires the group
integration formula~\cite{Creutz:1984mg},
\begin{eqnarray}
\int d\Omega \, \Omega_{ij} \Omega_{kl}^* = \frac{1}{N_c} \delta_{ik} \delta_{jl}  
\label{eq:7.4}
\end{eqnarray} 
whence one gets for the constant Polyakov loop case 
\begin{eqnarray}
\int d\Omega \, \tr_c \Omega \,  \tr_c \Omega^{-1} =1  
\label{eq:7.5}
\end{eqnarray}
Note that the effect of ignoring the Polyakov loop (i.e., setting
$\Omega=1)$ promotes this result by two orders in $N_c$. In this model
the average over pion fields is trivial since the vacuum energy
density does not depend on $U$ at the one quark loop level. Neglecting
momentarily the gluonic corrections $\Gamma_G$, using the previous
formulas and (\ref{eq:O_pnjl}) we get the leading order result
\begin{eqnarray}
L = \left\langle\frac{1}{N_c} \tr_c \Omega\right\rangle =
-\frac{1}{N_c^2} \frac{B V}{T} e^{-M_V/2T} \,.
\label{eq:L_SQM_lowT}
\end{eqnarray}
Note that at this order the contribution from the denominator is
trivial.  As expected, triality is not preserved due to the presence
of dynamical quarks, but the relevant scale is the constituent quark
mass. In addition, note that since $B$ is proportional to $N_c$ there
is an extra $1/N_c$ suppression. So the Polyakov loop can be
effectively used as an order parameter. Actually, our calculation
suggests that a low temperature calculation of the Polyakov loop in
full QCD might provide a method of extracting a gauge invariant
constituent quark mass. Proceeding in a similar way from the
expression of the quark condensate (\ref{eq:A.18a}) we get the leading
order contribution
\begin{eqnarray}
\frac{\langle \bar q q \rangle^*}{\langle \bar q q \rangle} = 1 +
\frac{2BV}{N_c^2 T} e^{-(M_V + M_S) /2T} + \cdots \,.
\end{eqnarray} 
It is noteworthy that the thermal correction scales as $1/N_c$ ($B$
scales as $N_c$), as in the ChPT case. This again is not just a
consequence of triality, but requires the proper integration over the
Polyakov loop manifold. The presence of the (infinite) four-volume
factor $V/T$ has to do with our assumption on a constant Polyakov
loop. As we have argued in Section~\ref{sec:local}, one has indeed a
local Polyakov loop and the volume should be replaced according to the
rule in Eq.~(\ref{eq:rule-vol}) by an effective confinement-domain
volume.\footnote{For the expectation value of a local
observable ${\cal O}(\vec{x})$, points outside the volume $V$ are not
correlated and their contribution approximately cancels in numerator
and denominator.}

The first gluonic correction contributes in $L$ as
$e^{-(M_V+2m_G)/2T}$, and in the quark condensate as
$e^{-(M_V+M_S+2m_G)/2T}$.

\subsection{NJL model}

The previous computation can also be considered within the NJL
model. In this model the fermion determinant can be obtained by means
of a derivative expansion~\cite{Megias:2002vr,Megias:2003ui}. The
result will be presented in Ref.~\cite{Megias:2006prep}. Retaining
only the vacuum contribution, which coincides with the result given in
Eq.~(3) of \cite{Fukushima:2003fw}, we have
\begin{equation}
\Det(i{\bf D})_\Omega = 
\exp \left( i\int d^4x \, ({\cal L}_q(T=0) + {\cal L}_q(\Omega,T))\right) \,,
\label{eq:det_NJL}
\end{equation}
where ${\cal L}_q(T=0)$ is the zero temperature contribution. At low
temperature, the thermal correction reads
\begin{equation}
{\cal L}_q(\Omega,T) = N_f\sqrt{\frac{M^3
T^5}{2\pi^3}}e^{-M/T}\tr_c\,(\Omega + \Omega^{-1}) + \cdots \,.
\end{equation} 
Using the volume rule $\int d^4x\,{\cal L}_q \to (V/T){\cal L}_q$,
expanding Eq.~(\ref{eq:det_NJL}) in powers of ${\cal L}_q(\Omega,T)$
and considering the group integration formula as above, we get the
leading order result\footnote{Actually we find a negative value for
the SQM and positive for the NJL model. While based on color-charge
conjugation symmetry it can rigorously be shown that $L$ must be real
no proof exists to our knowledge that $L > 0$ at any temperature,
although lattice data~\cite{Kaczmarek:2005ui} favor the positive
case.}
\begin{equation}
L = \left\langle\frac{1}{N_c} \tr_c \Omega\right\rangle =
\frac{N_f}{N_c} \frac{V}{T} \sqrt{\frac{M^3 T^5}{2\pi^3}} e^{-M/T} \,.
\label{eq:L_NJL_lowT}
\end{equation}
(Since the NJL bosonization term in (\ref{eq:det_b_Q}) cancels in the
calculation of observables, it needs not be included in this
calculation.  Also the gluonic corrections have been omitted. Their
effect is discussed below).

For the quark condensate we take into account the result similar to
Eq.~(\ref{eq:cond}) but replacing $(-1)^n$ with $(-\Omega)^n$,
corresponding to the quark condensate for fixed Polyakov loop. Thus,
including the leading fermion determinant contribution, using
(\ref{eq:7.4}), and taking into account that $\langle \tr_c
\Omega\rangle = \langle \tr_c \Omega^{-1}\rangle$, we get for the
single flavor condensate
\begin{equation}
\langle \bar q q\rangle^* = \langle \bar q q\rangle + \frac{ N_f 
V}{\pi^3}(M T)^3 e^{-2M/T} \,.
\label{eq:qq_NJL_lowT}
\end{equation}
Note that the $N_f$ factor comes from the fermion determinant.  As in
the spectral quark model, the first gluonic correction contributes in
$L$ with $e^{-(M+m_G)/T}$, and in the quark condensate
with~$e^{-(2M+m_G)/T}$.

As we see, beyond the quenched approximation the Polyakov cooling
persists although is a bit less effective as in the quenched case, and
for instance the temperature dependence of the low energy constants of
the tree level chiral effective Lagrangian becomes $ L_i^* - L_i
\stackrel{\rm Low \; T} \sim e^{- M_V /T} $~\cite{Megias:2006prep}.

Finally, on top of this one must include higher quark loops, or
equivalently mesonic excitations, from which the pions are the
dominant ones. They yield exactly the results of
ChPT~\cite{Florkowski:1996wf} for the chiral condensate $\langle \bar
q q \rangle $ and for the would-be Goldstone bosons, pions dominate at
low temperatures.  Thus, we see that when suitably coupled to chiral
quark models the Polyakov loop provides a quite natural explanation of
results found long ago on purely hadronic grounds~\cite{Gasser:1986vb}
as a direct consequence of the genuinely non-perturbative finite
temperature gluonic effects. The expected leading correction effect on
the Polyakov loop is also an additional exponential suppression ${\cal
O} (e^{-m_\pi /T})$.

\section{Implications for the phase transition} 
\label{sec:phase-tran}

The inclusion of the Polyakov loop has the consequence that one
changes the one quark state Boltzmann factor $N_c e^{-M/T} $ into $
\langle \tr_c \Omega \rangle $ at low temperatures. In the quenched
approximation one has $\langle \tr_c \Omega \rangle =0$, whereas the
first non-vanishing contribution stemming from the Dirac sea behaves
as $\langle \tr_c \Omega \rangle \sim e^{-M/T}$ due to the explicit
breaking of the center symmetry induced by the fermion determinant.
Likewise, for the quark condensate $\langle \bar q q \rangle $ the
finite temperature correction changes $N_c e^{-M/T} \to e^{-2 M/T} $
after the Polyakov loop integration is considered. Taking into account
the large number of approximations and possible sources of corrections
it is difficult to assess the accuracy of these Polyakov Chiral Quark
Models, in spite of the phenomenological success achieved in
Refs.~\cite{Fukushima:2003fw,Ratti:2005jh} within the mean field
approach. Nevertheless, it is tempting to see how these results may be
modified not only at low temperatures but also in the region around
the phase transition when the proper quantum and local nature of the
Polyakov loop is considered. This requires going beyond low
temperature truncations like (\ref{eq:L_NJL_lowT}) and
(\ref{eq:qq_NJL_lowT}).  Clearly a proper description would demand a
good knowledge of the Polyakov loop distribution as a function of the
temperature.  Unfortunately, such a distribution is poorly known and
lattice simulations are not designed to extract it, since a subtle
renormalization issue is involved
\cite{Kaczmarek:2002mc,Dumitru:2003hp}.  As a first step to
investigate the phase transition in the Polyakov chiral quark model
beyond the mean field approximation we just take the strong coupling
model for the gluonic action of (\ref{eq:3.8}).  Due to the rather
large exponential suppression, this ansatz has the virtue of reducing
to the Haar measure in the low temperature regime, and as a
consequence the vanishing of the adjoint Polyakov loop expectation
value observed in lattice calculations \cite{Dumitru:2003hp}
follows. In our view this is a compelling reason to go beyond mean
field by integrating over Polyakov loops. However, such a distribution
preserves center symmetry and would not generate a phase transition
per se in gluodynamics.  This is unlike the mean field approximation
where the action is minimized by center symmetry breaking
configurations. As discussed before a side product of this
approximation is to miss the fluctuations and also to introduce an
explicit coordinate dependence in the gauge group. In our model the
breaking of the center symmetry is attributed only to quarks. As we
will see this explicit breaking is rather large precisely due to the
simultaneous restoration of the chiral symmetry, since the constituent
quark mass drops to zero. The qualitative agreement with lattice
calculations in full QCD suggests that an important part of the
physics has been retained by the model, leaving room for improvement
in the Polyakov loop distribution.

We will present calculations only for the NJL model. In practice, we
use (\ref{eq:O_pnjl}), where the fermion determinant corresponds to
Eq.~(3) of \cite{Fukushima:2003fw} plus the volume rule
(\ref{eq:rule-vol}). The Polyakov loop integration is carried out
numerically. Due to gauge invariance the Polyakov loop dependence is
through its eigenvalues, and thus one may use the marginal
distribution of eigenvalues (\ref{eq:average1}), which for $N_c=3$
amounts to two independent integration variables. Full details are
given in appendix \ref{sec:njl_app}.

\begin{figure}[ttt]
\begin{center}
\epsfig{figure=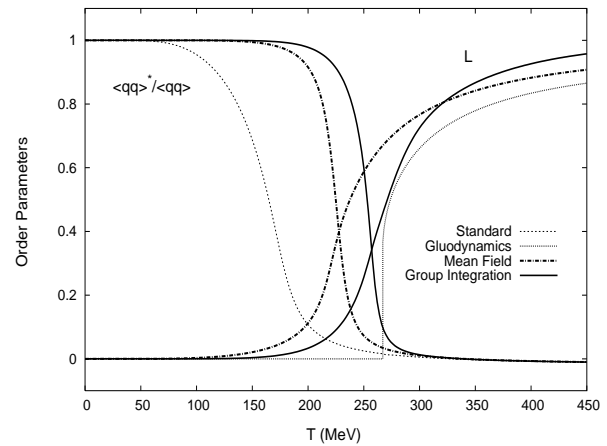,height=6cm,width=8cm}
\end{center}
\caption{Temperature dependence of the chiral condensate $\langle \bar
q q \rangle $ and Polyakov loop expectation value $ L= \langle \tr_c
\Omega \rangle /N_c $ in relative units. The standard result for
$\langle \bar q q\rangle^*$ corresponds to the pure NJL model uncoupled
to the Polyakov loop. The result of $L$ for gluodynamics within the
strong coupling expansion is also displayed. We compare the mean field
approach of Ref.~\cite{Fukushima:2003fw} where the Polyakov loop is
classical and coupled to the quarks, with the integration over the
Polyakov loop $\Omega$.}
\label{fig:phase_transition}
\end{figure}

\begin{figure}[ttt]
\begin{center}
\epsfig{figure=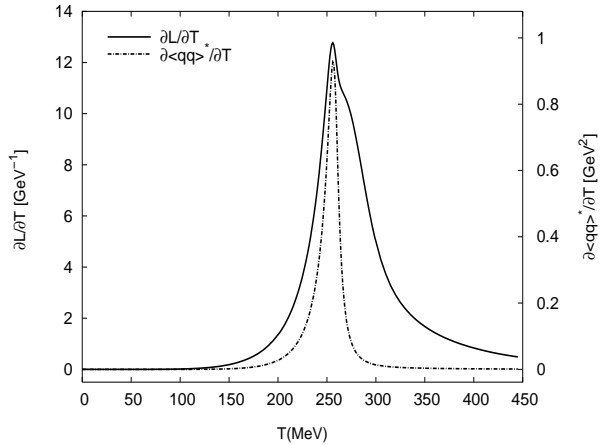,height=6cm,width=8cm}
\end{center}
\caption{Temperature dependence of $\partial\langle \bar q q
\rangle^*/\partial T $ and $ \partial L/\partial T$ in the NJL model
when the integration over the Polyakov loop~$\Omega$ is carried out.}
\label{fig:susc}
\end{figure}

In Fig.~\ref{fig:phase_transition} we show the effect on both the
chiral condensate $\langle \bar q q \rangle $ and Polyakov loop
expectation value $ L= \langle \tr_c \Omega \rangle /N_c $ within
several schemes. In all cases we always minimize with respect to the
quark mass and use $\rho(\Omega)$ in (\ref{eq:3.8}) for all
temperatures. We compare the standard NJL model with no Polyakov loop
with the mean field calculation of Ref.~\cite{Fukushima:2003fw}, which
corresponds to minimize the vacuum energy as a function of the
constituent mass and a given choice of the Polyakov loop matrix. We
also compare with the result one obtains by integrating in the
Polyakov loop instead and minimizing with respect to the quark mass
afterwards.  We work in these calculations with the NJL model with
2-flavor, $N_f=2$, and consider for the current quark mass matrix
$\hat M_0=\diag(m_u,m_d)$ the isospin-symmetric limit with $m_u = m_d
\equiv m_q = 5.5\,\text{MeV}$. The zero temperature part of the
effective action of Eq.~(\ref{eq:eff_ac_njl}) is regulated by the
Pauli-Villars method, with the cut-off $\Lambda_{\text PV}=
828\,\text{MeV}$, corresponding to a constituent quark mass
$M=300\,\text{MeV}$. The coupling is $G=13.13\,\text{GeV}^{-2}$, which is
obtained from the gap equation~(\ref{eq:gap_eq}). These parameters
reproduce the empirical values of the pion weak-decay constant and the
quark condensate at zero temperature. Aspects of locality have been
considered in the treatment of the NJL model with the integration in
the Polyakov loop, by introducing the volume rule~(\ref{eq:rule-vol}),
where the string tension has been fixed to its zero temperature
value~$\sigma=(425 \, \text{MeV})^2$. It is also displayed in the
figure the expectation value $L$ in gluodynamics within the model of
Eq.~(\ref{eq:G_potential_sce}) in the mean field approximation, which
leads to a first order phase transition at $T_D=270\,\text{MeV}$. As
we see the net effect of the Polyakov loop integration is to displace
the transition temperature to somewhat higher values.  So, the method
based on the integration provides an effective cooling at higher
temperatures for fixed parameters. As we can see in
Fig.~\ref{fig:susc}, the crossover transitions for the chiral
condensate~$\langle \bar q q\rangle$ and for the Polyakov loop
expectation value~$L$ coincide at the value~$T_c \simeq 256
\,\text{MeV}$.

\begin{figure}[ttt]
\begin{center}
\epsfig{figure=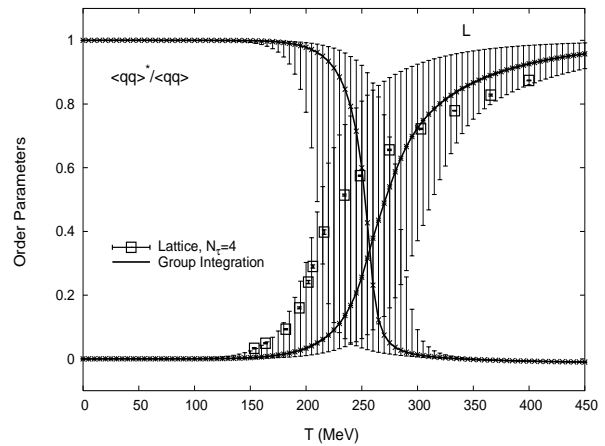,height=6cm,width=8cm}
\end{center}
\caption{Temperature dependence of the chiral condensate $\langle \bar
q q \rangle $ and Polyakov loop expectation value $ L= \langle \tr_c
\Omega \rangle /N_c $ in relative units, in the NJL model when the
integration over the Polyakov loop $\Omega$ is carried out. The error
bands are associated to an uncertainty in the string tension of
$\sigma = 0.181 \pm 0.085\,{\text GeV}^2$. We compare with lattice
data correspond to 2-flavor QCD, taken from \cite{Kaczmarek:2005ui}.}
\label{fig:phase_transition2}
\end{figure}

We have checked that a temperature dependence of the string tension
may accommodate the unquenched lattice
results~\cite{Kaczmarek:2005ui}, as we can see in
Fig.~\ref{fig:phase_transition2}. This provides a range of string
tensions $\sigma = 0.181 \pm 0.085 \,\text{GeV}^2$ with somehow
account for an estimate of the uncertainty in the present model. In
Fig.~\ref{fig:phase_transition2} the error band associated to such an
uncertainty reflects a critical temperature of about $T_D = 250 \pm 50
\,\text{MeV}$. This is compatible with the large rescaling advocated
in Ref.~\cite{Ratti:2005jh}. At present, and taking into account the
many possible sources of corrections to our calculations we do not see
how more accurate predictions could reliably be made in the context of
Polyakov-Chiral Quark Models. Nevertheless the semiquantitative
success indicates that essential features for the center symmetry
breaking phase transition are encapsulated by these models, and
further attempts along these lines should be striven. Nevertheless, it
should be reminded that although the breaking of the center symmetry
in this model is only attributed to the presence of quarks, one also
has a contribution from gluons. In this regard let us mention that
ignoring the exponentially suppressed gluon action (\ref{eq:3.8}) in
the averaging has almost no effect below the phase transition and
shifts up the transition temperature by about $ 30\, \text{MeV} $, a
value within our error estimate. Given the importance of quarks in the
phase transition one may wonder if the temperature dependent volume
enhances the breaking of the center symmetry. In fact, the volume at
the transition temperature is roughly equal to gluon volume $a^3$ in
(\ref{eq:G_potential_sce}). At low temperatures the exponential
suppression dominates in the Polyakov loop expectation value where the
volume appears as a harmless prefactor, see
e.g. (\ref{eq:L_NJL_lowT}). The effect of replacing the temperature
dependent volume by a constant one can be seen in
Fig. \ref{fig:volume}. Again changes are within our expected
uncertainties.

\begin{figure*}[tbc]
\begin{center}
\epsfig{figure=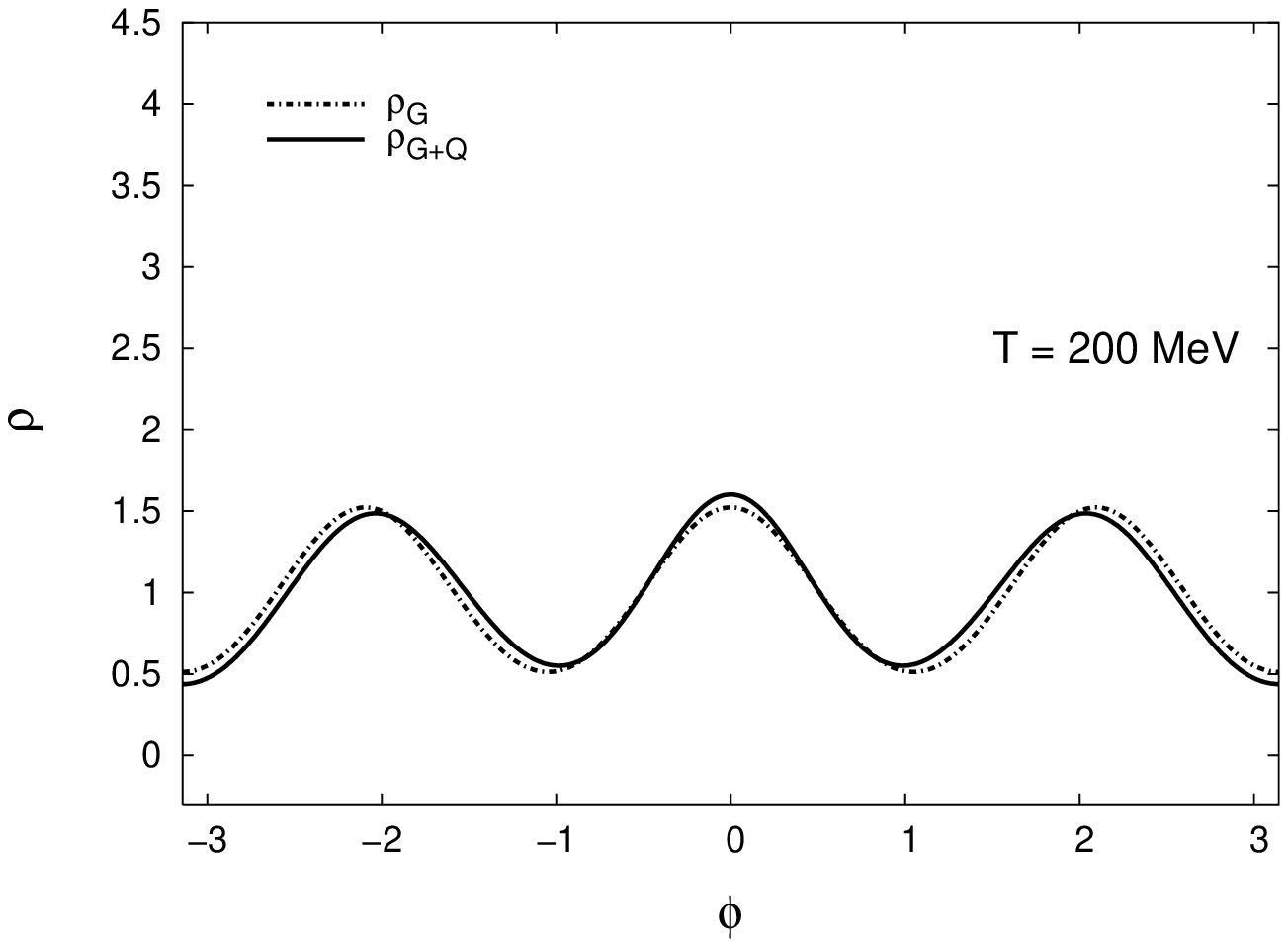,height=5cm,width=5cm}
\epsfig{figure=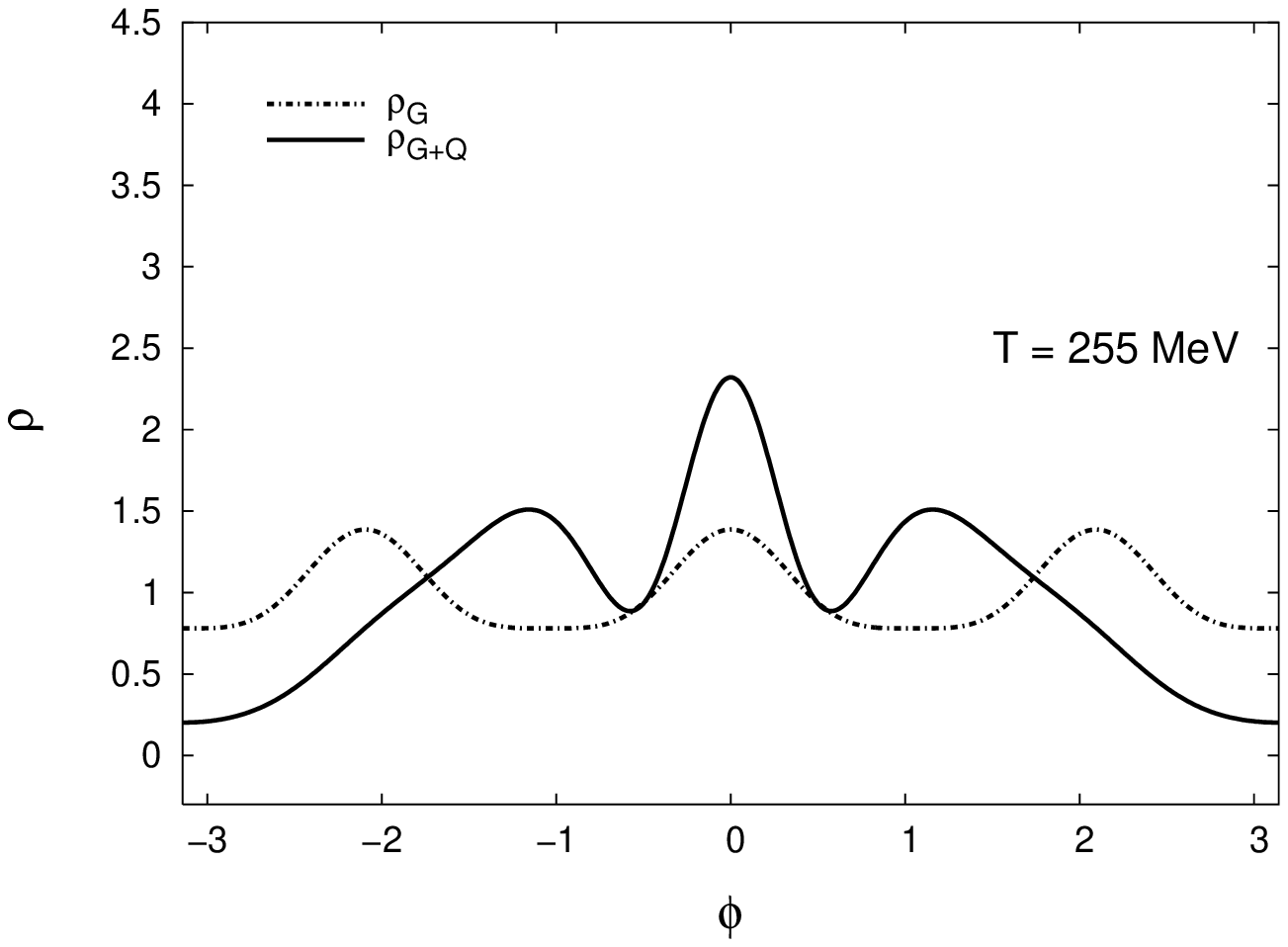,height=5cm,width=5cm}
\epsfig{figure=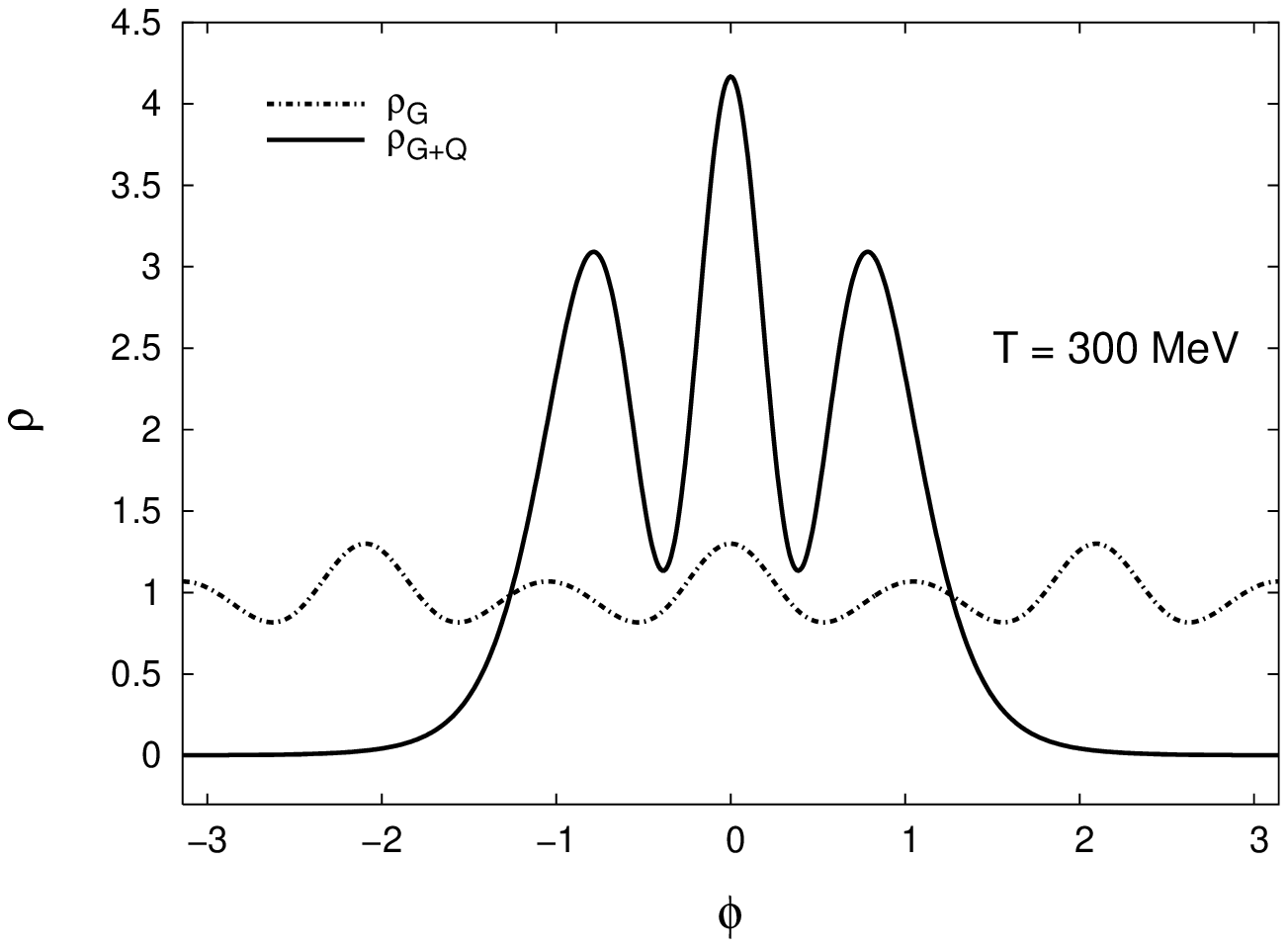,height=5cm,width=5cm} 
\end{center} 
\caption{Temperature dependence of the one-angle Polyakov loop
distribution $\hat\rho(\phi)$ of (\ref{eq:rhoG_rhoQ}) as a function of
the angle. Dash-dotted lines: quenched result ($\rho_G$ is included,
$\rho_Q$ is not). Center symmetry is preserved. Solid lines:
unquenched result (both factors $\rho_G$ and $\rho_Q$ are included) in
the NJL model. Center symmetry is explicitly broken. Three
temperatures nearby the transition (255$\,$MeV) are considered. For
convenience all distributions have been normalized to unity.}
\label{fig:polyakov_prob}
\end{figure*}

As we have argued the expectation value of the Polyakov loop is rather
small at temperatures well below the phase transition. The difference
between the mean field and the direct integration can be best
quantified at the level of the fluctuations. While at the mean field
level the probability of finding a given Polyakov loop would be a delta
function, one expects a spreading of such probability due to quantum
effects. For $N_c=3$ the Polyakov loop contains two independent
variables, which correspond to gluon fields in temperature units.
\begin{eqnarray}
\Omega = \diag \left( e^{i \phi_1} , e^{i \phi_2} , e^{-i (\phi_1+\phi_2) }\right) 
\label{eq:Omega_param}
\end{eqnarray} 
The joint distribution $\rho(\phi_1,\phi_2)$ can
be factorized as a product of the purely gluonic and the
quark determinant contributions (see appendix \ref{sec:njl_app})
\begin{eqnarray} 
\rho(\phi_1, \phi_2 ) = \rho_G (\phi_1, \phi_2 ) \rho_Q
(\phi_1, \phi_2 )
\end{eqnarray}
echoing the effective action displayed in Eq.~(\ref{eq:Z_pnjl}) in
Euclidean space. Note that $\rho(\phi_1, \phi_2 )$ is not normalized
to unity, instead its integral gives the full partition function (see
appendix \ref{sec:njl_app}). As noted in Sect.~\ref{sec:dyn-pol} by
gauge invariance the distribution is invariant under permutation of
the three angles $\phi_1$, $\phi_2$ and $\phi_3=-\phi_1-\phi_2$. The
use of such a symmetry is that the trace of any arbitrary function of
the Polyakov loop $ f(\Omega) $ (a one-body operator) can be averaged
over the group by integrating out one angle,
Eq.~(\ref{eq:one-body}). Thus one obtains an equivalent one-body
distribution as
\begin{eqnarray}
\widehat\rho(\phi) \propto \frac1{2\pi} \int_{-\pi}^\pi d \phi^\prime \rho_G ( \phi,
\phi^\prime ) \rho_Q ( \phi, \phi^\prime) \,.
\label{eq:rhoG_rhoQ}
\end{eqnarray} 
It is interesting to compare how this distribution evolves across the
phase transition, and to look for the effects generated explicitly by
the fermion determinant.  In Fig.~\ref{fig:polyakov_prob} we present
such a comparison.  Below the phase transition, and as already
advanced in Sect.~\ref{sec:dyn-pol}, the weighting function presents
three maxima at equidistant values, as required by the center
symmetry. In this case the quark determinant plays a negligible role,
although a tiny, indeed exponentially small, center symmetry breaking
can be observed. As we see there appears an interesting concentration
of angles in the region around the origin as the phase transition
takes place. The quarks are very effective suppressing contributions
not near $\Omega=1$. As a consequence the lack of the spontaneous
breaking of the center symmetry in (\ref{eq:3.8}) becomes not very
relevant for temperatures above the transition.

\begin{figure}[ttt]
\begin{center}
\epsfig{figure=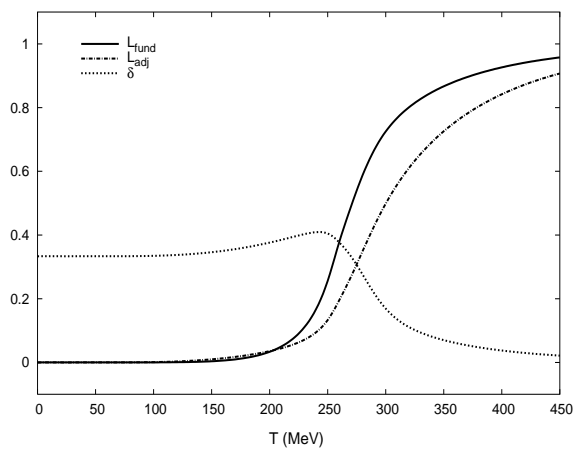,height=6cm,width=8cm}
\end{center}
\caption{Temperature dependence of the Polyakov loop expectation value
 in the fundamental, $ \langle \tr_c \,\Omega \rangle /N_c $, and adjoint,
$\langle \widehat {\rm tr}_c \, \widehat \Omega \rangle /(N_c^2-1) $,
representations and the total fluctuation $\delta$ of the Polyakov
loop, in the NJL model when the integration over the Polyakov loop is
carried out.}
\label{fig:fluctuation}
\end{figure}

A further trace of fluctuations can be seen by considering higher
group representations of the Polyakov loop. In
Fig.~\ref{fig:fluctuation} we also show the expectation value of the
Polyakov loop in the adjoint representation, $\langle \widehat {\rm
tr}_c \, \widehat \Omega \rangle /(N_c^2-1) $. According to the
lattice results of the matrix model in Ref.~\cite{Dumitru:2003hp} one
has a vanishing expectation below the phase transition. As we have
argued above, this feature is not preserved at the mean field level,
where a non-vanishing value $-1/(N_c^2-1)$ is obtained instead (see
Eq.~(\ref{eq:pol-adjoint}) for the case $n=1$). Considering the
Polyakov loop integration, as we do, complies with the lattice
expectations and indicates that further developments should consider
these constraints. The full fluctuation of the Polyakov loop,
$\delta$, is defined by
\begin{eqnarray}
\delta^2 &\equiv& \left(\langle |\tr_c\Omega |^2\rangle 
- \langle \tr_c \Omega \rangle^2 \right)/N_c^2 \,, \nonumber \\
&=& \left(1+ \langle \widehat\tr_c \widehat\Omega\rangle - 
\langle  \tr_c \Omega\rangle^2 \right)/N_c^2  \,.
\label{eq:fluctuation_L}
\end{eqnarray}
The fluctuation is also shown in Fig.~\ref{fig:fluctuation}. $\delta$
goes to zero in the large~$T$ regime, and this is compatible with the
fact that the one-body distribution~$\widehat\rho(\phi)$ tends to
concentrate near $\phi=0$ as the temperature increases. In the second
equality of Eq.~(\ref{eq:fluctuation_L}) we have used the
identity~(\ref{eq:pol-adjoint}) with $n=1$.

\section{conclusions}
\label{sec:concl} 

In the present work we have discussed how the pro\-blem of
conventional chiral quark models at finite temperature may be overcome
by introducing the Polyakov loop. In order to maintain gauge
invariance at finite temperature some non-perturbative explicit
gluonic degrees of freedom must be kept. In practice, and in particular
gauges such as the Polyakov gauge, the approach corresponds to treat
the $A_0$ component of the gluon field as a color dependent chemical
potential in the quark propagator. This introduces, however, a color
source which generates any possible color non-singlet states, calling
for a projection onto the physical color singlet states, or
equivalently evaluating the path integral over the $A_0$ field in a
gauge invariant fashion. As such, the average includes both the gluon
action and the quark determinant. Models for the gluonic part have
been discussed on the light of pure gluodynamics results on the
lattice.  The net result is that, contrary to standard chiral quark
model calculations at finite temperature, no single quark excitations
are allowed in physical observables.  More generally, the leading
thermal corrections at the one quark loop level start only at
temperatures near the deconfinement transition. Given the fact that
this strong suppression effect is triggered by a group averaging of
Polyakov loops we have named this effect Polyakov cooling of the quark
excitations. Thus, and to a very good approximation, we do not expect
any important finite temperature effect on quark observables below the
deconfinement transition. In particular the chiral symmetry breaking
transition cannot occur before the deconfinement transition. In such a
situation the biggest change of observables such as the quark
condensate should come from pseudoscalar loops at low temperatures and
perhaps other higher meson resonances at intermediate
temperatures. This is precisely what one expects from ChPT or
unitarized approaches thereof which effectively include these loops on
resonances. It is rewarding to see how in practice the apparent
contradiction between chiral quark models and ChPT in the intermediate
temperature region may be resolved by a judicious implementation of
the Polyakov loop.

The extrapolation of these ideas to the phase transition is
straightforward but more ingredients are needed. As an illustration we
have investigated in a model the kind of effects one might expect from
such a schematic Polyakov-Chiral Quark Model when both the quantum and
local nature of the Polyakov loop are taken into account. Several
interesting features arise out of such an investigation. At low
temperatures the Polyakov loop is suppressed exponentially in the
constituent quark mass suggesting that eventually more accurate
lattice measurements might provide a method to extract the constituent
quark mass in a gauge invariant fashion. According to our analysis
corrections to this leading behavior are provided by pion loops. It
would be extremely helpful to find a general theoretical setup where
these chiral corrections might be reliably computed. Moreover, we find
that the explicit breaking of the center symmetry due to dynamical
quarks at low temperature is $1/N_c $ suppressed. This is a
direct consequence of averaging over gauge field configurations and
confirms the current usage of the Polyakov loop as an order parameter
in the unquenched case. On the light of the present findings one might
conjecture that in the large $N_c$ limit the Polyakov loop becomes a
true order parameter of full QCD.

Another feature we find is that the contribution of the gluon dynamics
below the phase transition does not seem to be crucial. This is
welcome since this is precisely the region where least information can
be deduced from lattice simulations besides the known preservation of
the center symmetry. Nevertheless, it would rather interesting for our
understanding of the low temperature gluon dynamics to compute
directly from the lattice the Polyakov loop probability distribution.
From our results we deduce that although the qualitative features
observed in more simplified treatments are confirmed by calculations,
one might expect large uncertainties in the determination of critical
parameters, such as the critical temperature. Our estimate is $T_D =
250 \pm 50 {\rm MeV}$ for $N_f=2$. Even given these large
uncertainties, the very fact that a crossover between chiral symmetry
restoration and center symmetry breaking takes place in the bulk part
of the expected lattice QCD simulations with a minimal number of
parameters is very encouraging and motivates that further studies
along these lines should be pursued. 

Finally, a more intriguing aspect regards what kind of model
independent information could be inferred out of these models, where
quarks and Polyakov loops are coupled, in the regime around the phase
transition. For instance, the low temperature behavior of the chiral
condensate can be described using Chiral Perturbation Theory in terms
of the zero temperature chiral condensate with no explicit reference
to the underlying quark and gluonic degrees of freedom due to the
dominance of pionic fluctuations. Given the fact that the Polyakov
loop is a gauge invariant object which vanishes at zero temperature,
it would be extremely helpful to isolate what physical states could
equivalently describe such an observable and what specific zero
temperature QCD operators drive its low temperature behavior.

\begin{acknowledgments}
We thank W. Broniowski for discussions. This work is supported in part
by funds provided by the Spanish DGI and FEDER founds with grant
no. FIS2005-00810, Junta de Andaluc\'{\i}a grant no. FM-225 and
EURIDICE grant number HPRN-CT-2003-00311.
\end{acknowledgments}

\appendix 

\section{The Spectral Quark Model at Finite Temperature} 
\label{sec:sqm} 

In this Appendix we show how calculations for the spectral quark model
introduced in Refs.~\cite{RuizArriola:2001rr, RuizArriola:2003bs,
RuizArriola:2003wi, Megias:2004uj} can be extended at finite
temperature.

\subsection{The spectral quark model at zero temperature}

In the spectral quark model~
\cite{RuizArriola:2001rr,RuizArriola:2003bs} the quark propagator is
written in the generalized Lehmann form
\begin{eqnarray}
S(k) &=& \int_C d \omega { \rho( \omega ) \over \k - \omega
} = \k A(k^2 ) + B(k^2) 
\nonumber \\
 &=& \frac{Z(k^2)}{\k - M(k^2)}
\label{eq:lehmann} 
\end{eqnarray}
where $\rho(\omega)$ is the, generally complex, spectral function and
$C$ denotes a contour in the complex $\omega$ plane. $M(k^2)$ is the
self energy and $Z(k^2)$ is the wave function renormalization. As
discussed already in Ref.~\cite{RuizArriola:2003bs} the proper
normalization and the conditions of finiteness of hadronic observables
are achieved by requesting an infinite set of {\em spectral
conditions} for the moments of the quark spectral function,
$\rho(\omega)$, namely
\begin{eqnarray}
\rho_0 & \equiv & \int d\omega \rho(\omega ) = 1, \label{rho0} \\
\rho_n & \equiv & \int d\omega \omega^n \rho(\omega) = 0, 
\quad \text{for~~}  n=1,2,3,\ldots \label{rhon}
\end{eqnarray} 
Physical observables are proportional the zeroth and the inverse
moments,
\begin{eqnarray}
\rho_{-n} & \equiv & \int d\omega \omega^{-n} \rho(\omega), \quad
\text{for~~}  n=0,1, 2,\ldots \label{rhoinv}
\end{eqnarray} 
as well as to the {\em ``log moments''},
\begin{eqnarray}
\rho^\prime_n &\equiv&  \int d\omega \log(\omega^2) \omega^n
\rho(\omega), \;\;\; \text{for} \; n=1,2,3,\ldots
\label{rholog}
\end{eqnarray}
Obviously, when an observable is proportional to the dimensionless
$\rho_0=1$ the result does not depend explicitly on the
regularization. The spectral conditions (\ref{rhon}) remove the
dependence on a scale $\mu$ in (\ref{rholog}), thus guaran\-tying the
absence of any dimensional transmutation. No standard requirement of
positivity for the spectral strength, $\rho(\omega)$, is made. The
spectral regularization is a physical regularization in the sense that
it provides a high energy suppression in one quark loop amplitudes and
will not be removed at the end of the calculation. Using the methods
of Ref.~\cite{Megias:2004uj} one finds at zero tempera\-ture, $T=0$, and
trivial Polyakov loop $\Omega=1$, and in the chiral limit, $m_\pi=0$,
the following results for the pion weak decay constant, $f_\pi$, the
single flavor condensate $\langle \bar q q \rangle $, the
energy-momentum tensor $\theta^{\mu\nu}$ and the anomalous $\pi^0 \to
2 \gamma $ amplitude
\begin{eqnarray}
f_\pi^2 &=& 4iN_c \int \frac{d^4 k}{(2\pi)^4} \left[ k^2 A(k^2)
\right]^\prime 
\\ 
N_f \langle \bar q q \rangle &=& - 4iN_c N_f \int \frac{d^4
k}{(2\pi)^4} B(k^2)
\\ 
\langle \theta^{\mu \nu} \rangle &=& -4i N_c N_f \int \frac{d^4
k}{(2\pi)^4} \left[ k^\mu k^\nu A (k^2)-g^{\mu \nu} \right]
\\ 
F_{\pi \gamma \gamma } &=& i \frac{4}{N_c f_\pi} \int \frac{d^4
k}{(2\pi)^4} \left[ k^2 A(k^2) \right]^{\prime\prime}  
\end{eqnarray} 
The vacuum energy density is defined by $ \epsilon_V = \frac14 \langle
\theta^\mu_\mu \rangle = -B $ with $B$ the bag constant. In the meson
dominance (MD) version of the SQM one obtains for the even an odd
components of the spectral function
\begin{eqnarray}
\rho_V(\omega) &=& \frac12 \left[ {\rho(\omega)}+ \rho(-\omega)
\right] \nonumber \\ &=& \frac{1}{2\pi
i}\frac{1}{\omega}\frac{1}{(1-4\omega^2/M_V^2)^{5/2}} \,, \\
\rho_S(\omega) &=& \frac12 \left[ {\rho(\omega)}- \rho(-\omega)
\right] \nonumber \\
&=&\frac{1}{2\pi
i}\frac{12\rho_3^\prime}{M_S^4(1-4\omega^2/M_S^2)^{5/2}} \,.
\end{eqnarray}
($M_V$ and $M_S$ are the vector and scalar meson masses, respectively)
and hence
\begin{eqnarray}
A(k^2) &=& \frac1{k^2} \left[ 1 - \frac1{(1-4k^2/M_V^2)^{5/2}} \right],
\nonumber \\ 
B(k^2) &=& \frac{48 \pi^2 \langle \bar q q\rangle }
{M_S^4 N_c (1-4k^2/M_S^2)^{5/2}} \label{AB}.
\end{eqnarray} 
Thus, 
\begin{eqnarray}
f_\pi^2 &=& \frac{M_V^2 N_c }{24\pi^2} \,,\\ 
\epsilon_V &=& -B = -\frac{M_V^4 N_f N_c }{192 \pi^2} = -\frac{N_f}8 f_\pi^2 M_V^2 \,.
\end{eqnarray} 
For three flavors one has $B = (0.2 {\rm GeV})^4 $ for $M_V = 770 {\rm
MeV}$.

\subsection{The spectral quark model at finite temperature and arbitrary 
Polyakov loop}

We introduce finite temperature and arbitrary Polyakov loop by using
the rule
\begin{eqnarray}
\int \frac{d k_0}{2\pi} F(  \vec k,k_0 ) \to i T
\sum_{n=-\infty}^\infty  F( \vec k,i \om_n )
\label{eq:A.15}
\end{eqnarray}
with $\om_n$ the fermionic Matsubara frequencies, $ \om_n = 2 \pi T (
n+1/2 + \hat \nu) $, shifted by the logarithm of the Polyakov loop $
\Omega = e^{i 2 \pi \hat \nu} $. In the meson dominance model, we use
the momentum space representation, evaluate first the three
dimensional integrals and finally the sums over the Matsubara
frequencies. In practice all sums appearing are of the form 
\begin{eqnarray}
S_l (M,T) &=& T \sum_{n=-\infty}^\infty \frac1{(M^2 + \om_n^2)^l}
\nonumber \\
&=& \frac{1}{(l-1)!} \left(-\frac{d}{dM^2}\right)^{l-1} S_1 (M,T) \,.
\end{eqnarray}
The basic sum is given by
\begin{equation}
S_1(M,T)= \frac1{2 M} \frac{\sinh(M / T )}{ \cos(2\pi \hat \nu)+
\cosh(M / T)} \,.
\label{eq:A.18}
\end{equation}
Using the relations in \cite{RuizArriola:2003bs} and the previous
formulas, we get
\begin{eqnarray}
\frac{\langle \bar q q \rangle^*}{\langle \bar q q \rangle} =
\frac1{N_c} \tr_c \left[\frac{\sinh(M_S/2T )}{ \cos(2\pi \hat
\nu)+ \cosh(M_S/2T)} \right]
\label{eq:A.18a}
\end{eqnarray} 
and, for the vacuum energy density 
\begin{eqnarray}
\frac{B^*}{B} =\frac1{N_c} \tr_c \left[\frac{\sinh(M_V/2T )}{
\cos(2\pi \hat \nu)+ \cosh(M_V/2T)} \right] \,.
\label{eq:A.19}
\end{eqnarray} 
Note that the relative temperature dependence of the bag constant and
the quark condensate are alike in the present model, if we also
interchange the vector and scalar masses. The $T$ and $\Omega$
dependence of $f_\pi^2$ is
\begin{widetext}
\begin{eqnarray}
\frac{f_\pi^*{}^2}{f_\pi^2} &=& \frac1{N_c} \tr_c \left[ \frac{ T
\sinh(M_V/ T) - M_V - \cos(2\pi \hat \nu) \left[ M_V \cosh(M_V /2T) -
2 T \sinh(M_V /2T) \right] }{ 2 T \left( \cos(2\pi \hat \nu)+
\cosh(M_V/2T) \right)^2} \right] \,.
\end{eqnarray} 
\end{widetext} 
To compute the group averages we observe that for a sum of the form of
Eq.~(\ref{eq:A.18}) we can make $ z= e^{i\phi} $ and $ t= e^{-M /T }
$. Expanding in powers of $t$ we get
\begin{eqnarray}
f(z) &=& \frac{1/t-t}{z+1/z+t+1/t} \nonumber \\ 
&=&   -1+\sum_{n=0}^\infty
(-t)^n \left( z^n + z^{-n} \right) \,,
\end{eqnarray}
and applying Eq.~(\ref{eq:p1})-(\ref{eq:p3}) for the average over the
SU($N_c$) Haar measure one has
\begin{eqnarray}
\left\langle \frac{1}{N_c} \tr_c f(z) \right\rangle = 1 -
\frac{2}{N_c} t^{N_c} \,.
\end{eqnarray}
Thus, undoing the change of variables we get  
\begin{equation}
\left\langle \frac{1}{N_c} \tr_c \frac{\sinh(M / T )}{ \cos(2\pi \hat \nu)
+ \cosh(M / T)}
\right\rangle = 1 - \frac{2}{N_c} e^{-N_c M/T } \,.
\end{equation}
Beyond the quenched approximation, integrals may be computed at low
temperatures using Eq.~(\ref{eq:7.5}) and generalizations thereof.

\section{Details on numerical group integration}
\label{sec:njl_app}

In this Appendix we give details relative to the calculation presented
in Sec.~\ref{sec:phase-tran} for the Nambu--Jona-Lasinio model.

The chiral condensate is obtained from maximization of the partition
function $Z$, Eq.~(\ref{eq:Z_pnjl}), with respect to the constituent
quark mass. To compute $Z$ we need to carry out first the color group
integration. To this end we consider the Polyakov gauge and
parameterize the SU(3) Polyakov loop matrix as in
Eq.~(\ref{eq:Omega_param}). The expression is
\begin{equation}  
Z = \int_{-\pi}^\pi \frac{d\phi_1}{2\pi} \frac{d\phi_2}{2\pi} 
\, \rho_G(\phi_1,\phi_2) \rho_Q(\phi_1,\phi_2) \,, 
\label{eq:Z_pnjl_param} 
\end{equation}
where we have separated a gluonic distribution
\begin{eqnarray}
D\Omega \, e^{i\Gamma_G[\Omega]} = \frac{d\phi_1}{2\pi} \frac{d\phi_2}{2\pi}  
\, \rho_G(\phi_1,\phi_2)
\end{eqnarray}
and a fermionic one
\begin{eqnarray}
e^{i \Gamma_Q[\Omega]} = \rho_Q(\phi_1,\phi_2) \,.
\end{eqnarray}
Note that in $\Gamma_Q[\Omega]$ we do not consider any dependence in the
mesonic $U$ fields, because we only retain the vacuum contribution. $\rho_G$
contains the Haar measure associated with the SU(3) group integration, as well
as the gluonic corrections given in Eq.~(\ref{eq:G_potential_sce}), i.e.
\begin{eqnarray}
\rho_G(\phi_1,\phi_2) &=& \frac{1}{6} \left(
27-|\tr_c\Omega|^4 + 8 \textrm{Re}((\tr_c\Omega)^3) - 18|\tr_c \Omega|^2
\right) \nonumber \\
&&\times\exp \left(
2(d-1) e^{-\sigma a/T} |\tr_c\Omega|^2
\right) \,,
\end{eqnarray}
and
\begin{equation}
\tr_c(\Omega)= e^{i\phi_1}+e^{i\phi_2}+e^{-i(\phi_1+\phi_2)} \,.
\end{equation}

\begin{figure}[ttt]
\begin{center}
\epsfig{figure=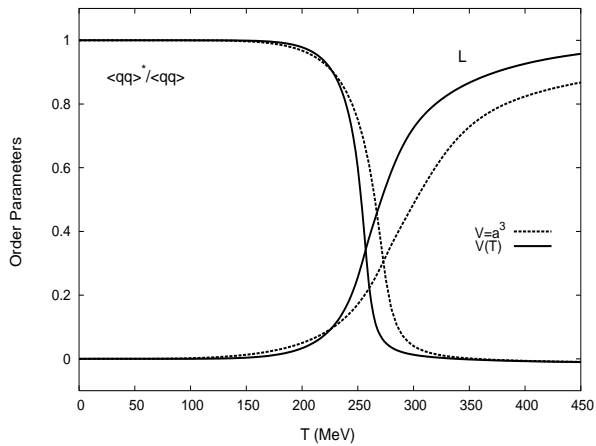,height=6cm,width=8cm}
\end{center}
\caption{Comparison of observables computed using a constant
correlation volume $a^3$, $a^{-1}=272\,$MeV (dashed line) versus the
temperature dependent volume $V$ of Eq.~(\ref{eq:rule-vol}) (solid
line).}
\label{fig:volume}
\end{figure}

The quark distribution $\rho_Q(\phi_1,\phi_2)$ follows from the vacuum
contribution in Eqs.~(\ref{eq:det_b_Q}) and (\ref{eq:det_NJL}). To
obtain $\Gamma_Q[\Omega]$ we use, passing over to Euclidean space,
\begin{eqnarray}
i\Gamma_Q[\Omega]&=&-\frac{V}{T}\Big({\cal L}_q(T=0) 
+ {\cal L}_q(T,\Omega)
\nonumber \\ &&
+\frac{1}{4G} \tr_f (M-\hat{M}_0)^2\Big) \,.
\end{eqnarray}
where the correlation volume $V$ is given in
(\ref{eq:rule-vol}). Moreover,
\begin{eqnarray}
{\cal L}_q(T=0)  &=&  -2 N_c N_f \int \frac{d^3
  k}{(2\pi)^3} E_k   \,,
\\
{\cal L}_q(T,\Omega) &=&  -2 T N_f \int \frac{d^3
  k}{(2\pi)^3} \bigg( 
 \tr_c \log \big[ 1+\Omega \, e^{-E_k/T}\big] 
\nonumber \\
&& +  \tr_c \log \big[ 1+\Omega^\dagger \,e^{-E_k/T}\big] 
\bigg)  \,.
\end{eqnarray}
Here $E_k = \sqrt{\mathbf{k}^2+M^2}$ is the energy of quasi-quarks and
the constituent quark mass is $M = m_q - G \langle
\bar{q}q\rangle$. For the zero temperature term we use the
Pauli-Villars regularization. After momentum integration we get
\begin{eqnarray}
{\cal L}_q(T=0)  &=&  - \frac{N_c N_f }{(4\pi)^2} 
\sum_i c_i(\Lambda_i^2+M^2)^2\log(\Lambda_i^2+M^2)
\nonumber \\
\end{eqnarray}
For the temperature dependent part, after momentum integration and
expanding the logarithm function, we obtain
\begin{eqnarray}
{\cal L}_q(T,\Omega) &=& \frac{N_f}{\pi^2} (M T)^2 \sum_{n=1}^\infty 
\frac{(-1)^n}{n^2}K_2(n M/T) \nonumber \\
&&\times\left( \tr_c (\Omega^n) + \tr_c (\Omega^{-n}) \right)
\end{eqnarray}

In numerical calculations we check for convergence in theses sums.
The expectation value of any observable is computed as in
Eq.~(\ref{eq:O_pnjl}). For any general function $f(\Omega)$, the group
averaging can be advantageously evaluated using
Eq.~(\ref{eq:one-body}), where the one-body distribution is given by
Eq.~(\ref{eq:rhoG_rhoQ}). It proves convenient to evaluate the double
integral as an iterated one, since with the exception of the adjoint
Polyakov loop, observables depend on one angle only. A comparison of
observables calculated using a constant correlation volume $a^3$,
$a^{-1}=272\,$MeV \cite{Fukushima:2003fw} and the temperature
dependent volume $V$ of Eq.~(\ref{eq:rule-vol}) is shown in
Fig.~\ref{fig:volume}.

%\bibliographystyle{h-physrev4}
%\bibliography{Refs1}

\end{document}